# The Dynamics and Scaling Laws of Planetary Dynamos Driven by Inertial Waves.

| | |
|---|---|
| Journal: | *Geophysical Journal International* |
| Manuscript ID: | GJI-14-0124 |
| Manuscript Type: | Research Paper |
| Date Submitted by the Author: | 13-Feb-2014 |
| Complete List of Authors: | Davidson, Peter; Cambridge University, Engineering |
| Keywords: | Dynamo: theories and simulations < GEOMAGNETISM and ELECTROMAGNETISM, Geomagnetic induction < GEOMAGNETISM and ELECTROMAGNETISM, Planetary interiors < PLANETS |
| | |

# The Dynamics and Scaling Laws of Planetary Dynamos Driven by Inertial Waves.

P. A. Davidson
Dept. Engineering, University of Cambridge, U.K.,

**Abstract**.
Great progress has been made in the numerical simulation of planetary dynamos, though these numerical experiments still operate in a regime very far from the planets. For example, it seems unlikely that viscous forces are at all significant in planetary interiors, yet some of the simulations display a significant dependence on viscosity, and indeed in some of the simulations the dynamo mechanism is itself viscously driven, taking the form of helical Ekman pumping within columnar convection rolls. Given the similarity of the external magnetic fields observed in the terrestrial planets and gas giants, and the extremely small value of the Ekman number in all such cases, it seems natural to suppose that the underlying dynamo mechanism in these planets is simple, robust, independent of viscosity and insensitive to mechanical boundary conditions. A key step to identifying this mechanism is to determine the source of helicity in planetary cores, which itself should be robust, independent of viscosity and insensitive to boundary conditions.

In this paper we explore the possibility that the helicity in the core of the Earth arises from the spontaneous emission of inertial waves, driven by the equatorial heat flux in the outer core. We also ask if a similar mechanism might operate in other planets, and perhaps act to supplement the helicity driven by Ekman pumping in the (viscous) numerical simulations. We demonstrate that such waves do indeed produce the required helicity distribution outside the tangent cylinder. Moreover, we show that these waves inevitably propagate along the axis of the columnar vortices, and indeed they are the very mechanism by which the columnar vortices form in the first place and the means by which the columns subsequently evolve. We also calculate the emf induced by such axially propagating inertial waves and show that, in principle, this emf is sufficient to support a self-sustaining dynamo of the $\alpha^2$ type. Finally, we derive the scaling laws for this kind of inertial-wave dynamo. We compare these predictions with the (imperfect) simulations, and also with what little we know about the Earth's core. The numerical experiments fall into two categories; the slowly rotating simulations which cannot sustain inertial waves at the small scales and the rapidly rotating (planet-like) ones which can. Our scaling laws are consistent with the latter class of simulations, and also with what we know about the Earth.

## 1. Introduction

### 1.1. The Need for a Simple, Robust Source of Helicity in Planetary Dynamos

Speculative models of a geodynamo driven by helical convection in the outer core have been around for over half a century (Parker, 1955). In recent years the numerical simulations of planetary dynamos have become sufficiently ambitious that they increasingly influence our thinking as to the dominant dynamics and field stretching mechanisms (Roberts & King, 2013, Olson, 2013). Of course, these simulations are overly viscous by a factor of around $\sim 10^9$, as measured by the Ekman number, $\mathrm{Ek} = \nu / \Omega R_C^2$, underpowered by a factor of around $\sim 100$, as measured by the Rayleigh number, and under rotate by a factor of around $\sim 10^3$, as measured by the Rossby number, $\mathrm{Ro} = |\mathbf{u}| / \Omega R_C$, where $\nu$ is the kinematic viscosity, $\Omega$ the planetary rotation rate, and $R_C$ the radius of the conducting core. (See, for example, the review by Christensen, 2011, for a comparison of the parameter regime captured by the simulations as opposed to the expected values in the planets.) Yet many of these simulations seem to yield plausible dynamos, with a dominant dipole structure exterior to the core which is more or less aligned with the rotation axis. Tentative confidence in these simulations has now grown to the extent that some researchers use them to establish dynamo scaling laws which are sometimes extrapolated to astrophysical objects (Christensen & Aubert, 2006, Christensen, 2010, Stelzer & Jackson, 2013, Davidson, 2013b). Moreover, a cartoon for planetary dynamo action has emerged from these numerical simulations which, given the plausibility of the simulations, has gained traction. In this cartoon the flow outside the tangent cylinder is dominated by columnar vortices aligned with the rotation axis, often in the form of alternating cyclones and anticyclones arranged around the solid inner core. Crucially, these columnar vortices are observed to act as conduits for helical motion, with the helicity outside the tangent cylinder, $h = \mathbf{u}\cdot\nabla\times\mathbf{u}$, predominantly positive in the south and negative in the north. It has long been known that such an antisymmetric helicity distribution can sustain a quasi-steady $\alpha^2$ dynamo of dipole structure (as discussed in, say, Moffatt, 1978), and so it has become popular to label the geodynamo as approximately $\alpha^2$, located outside the tangent cylinder, and driven by helical motion in the columnar convection cells (see, for example, Jones, 2011, or Roberts & King, 2013).

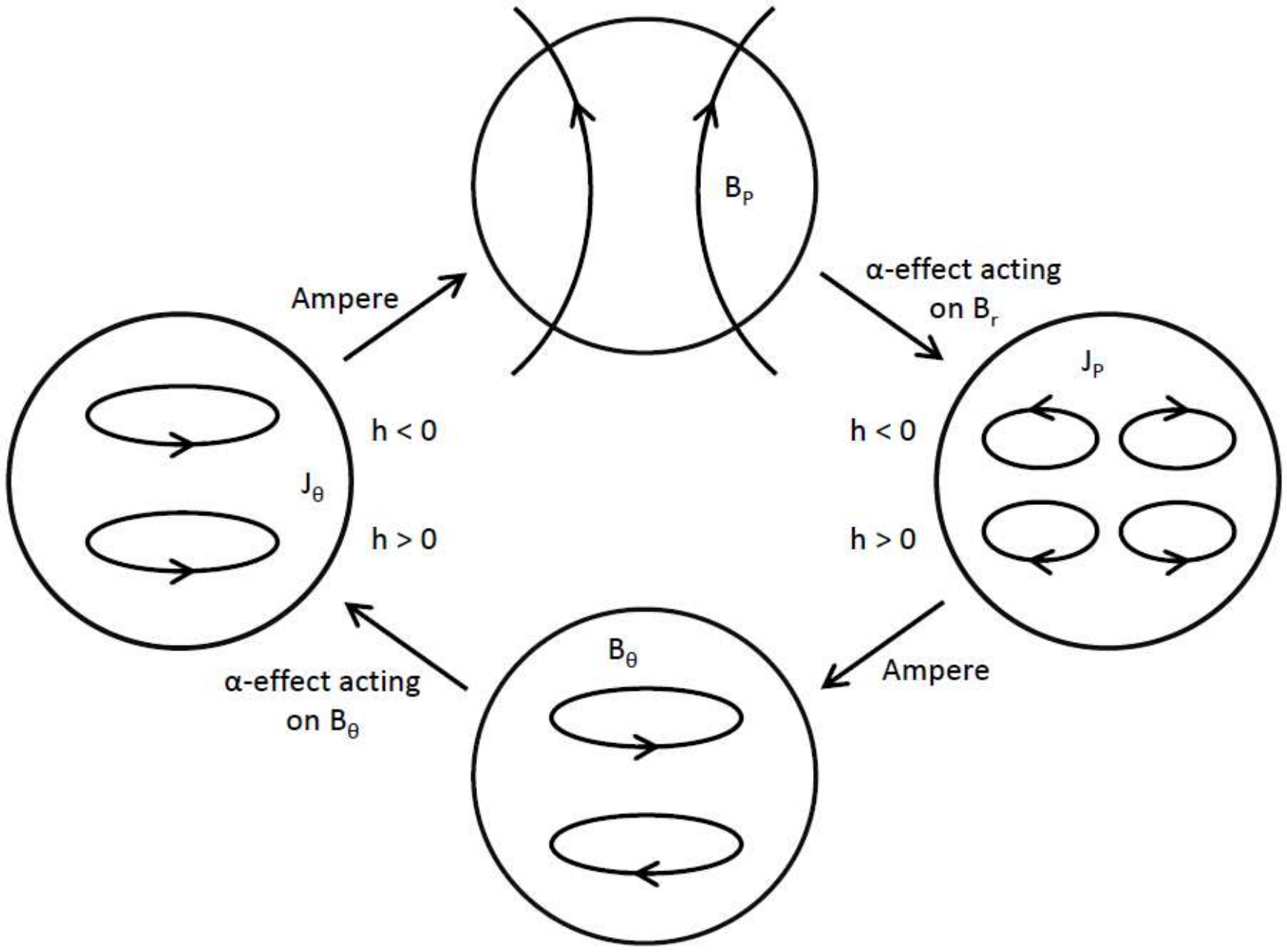

Figure 1. A classical $\alpha^2$ dynamo driven by helicity whose sign is antisymmetric about the equator. To focus thoughts, the dipole points to the north. When *h* is positive in the south and negative in the north we get $B_\theta$ to be positive in the north and negative in the south, which is more or less the distributions of *h* and $B_\theta$ seen outside the tangent cylinder in most of the numerical simulations.

This kind of classical $\alpha^2$ dynamo cycle is shown schematically in Figure 1, where we have chosen the signs of *h* in the north and the south to be the same as those observed outside the tangent cylinder in the numerical simulations. Here (and elsewhere in the paper) we use cylindrical polar coordinates $(r,\theta,z)$, $\mathbf{B}_p$ and $\mathbf{B}_\theta$ are the poloidal and azimuthal components of the axisymmetric part of the magnetic field, $\mathbf{J}_p$ is the poloidal current density induced by the $\alpha$-effect operating on the radial magnetic field, $B_r$, and $\mathbf{J}_\theta$ is the azimuthal current driven by the $\alpha$-effect operating on the azimuthal magnetic field, $B_\theta$. The $\alpha$-effect itself is assumed to be generated by helical motion within thin, axial columnar vortices which lifts and twists the radial and azimuthal components of the mean field, but does not act on the axial field (see §3). According to classical theory, to get a self-consistent dipolar dynamo with helicity antisymmetric about the equator, $\mathbf{B}_\theta$ must also be antisymmetric about the equator. In particular, with negative helicity in the north and positive helicity in the south, we require a mean azimuthal field outside the tangent cylinder which is positive in the north and negative in the south, which is more or less what is observed in most numerical simulations. (If the sign of the helicity is reversed, so the sign of the mean azimuthal field reverses, though the dynamo can still operate.)

Of course, the formalism of mean-field electrodynamics as applied to planetary cores has long since been questioned, as there is probably no clear separation of scales. However, the ideas and terminology of mean-field dynamos is still a useful way to categorize the results of the numerical simulations, and to capture the various dynamo mechanisms which manifest themselves, if only in a cartoon-like fashion.

Clearly, if we are to extend the dynamo cartoon suggested by the simulations to actual planets, it is crucial to identify what drives the helical flow in the convection cells and establish why this provides a pattern which is predominantly negative in the north and positive in the south. In the numerical simulations, which are of course much too viscous, the helicity is often attributed, at least in part, to viscous Ekman pumping driven by the interaction of the convection columns with the mantle (see, for example, Roberts & King, 2013). In short, the columnar vortices are seen as quasi-steady structures that span the core and grind away relentlessly on the mantel, inducing helical Ekman pumping. Certainly, such Ekman pumping does indeed produce the required antisymmetric distribution of helicity (negative in the north, positive in the south). It would seem, therefore, that we might have a zero-order model of the geodynamo.

There is a problem, however, with this cartoon. It is unlikely that the helicity in a planet would be generated by viscous boundary layers located on the mantle. There are a number of reasons for believing this. First, the observed properties of the dynamos operating in the gas giants are surprising similar to those in the terrestrial planets; they are predominantly dipolar with the magnetic axis roughly aligned with the rotation axis. Moreover, depending on how one normalises the dipole moment, $\mathbf{m}$, or equivalently the mean axial field strength in the core,

$$\bar{B}_z = \frac{2\mu}{3V_C}|\mathbf{m}|, \tag{1.1}$$

the induced magnetic field strengths look surprisingly similar across the planets, as shown by the last column in Table 1. (In equation 1.1, which may be found in, for example, Jackson, 1998, $V_C$ is the volume of the conducting core and $\mu$ the permeability of free space.) It seems plausible, therefore, that the underlying structure of the dynamos operating in the gas giants is not dissimilar to that in the Earth, yet there is no mantle in the gas giants on which Ekman layers can form. Second, it is striking that numerical simulations which employ slip boundary conditions on the mantle produce dynamos which are surprisingly similar to those with no-slip boundary conditions (see, Yadav et al, 2012), yet there are no Ekman layers in the former. Third, while it is clear that the columnar vortices can stretch from mantle to mantle in the mildly supercritical dynamo simulations, from which the popular dynamo cartoon has emerged, it is less clear that the majority of the columnar vortices extend across the entire core in the more strongly supercritical numerical simulations (see, for example, the images in Christensen & Wicht, 2007). Given that the Rayleigh number in the Earth is likely to be two orders of magnitude larger than that in the most strongly driven simulations to date, this becomes a major consideration when looking at the planets. Fourth, even if we believed that the columnar vortices spanned the core of the Earth, an estimated Ekman number of the order of $\mathrm{Ek} \sim 10^{-15}$ demands that standard viscous scaling theory yields an Ekman layer thickness, $\delta_E \sim (\mathrm{Ek})^{1/2} R_C$, of around 10cm and a convection column width, $\delta \sim (\mathrm{Ek})^{1/3} R_C$, of a few tens of metres. Such small values seem inconceivable in the outer core, yielding Ekman layers which are a tiny fraction of the height of the undulations on the mantle and inner core boundary, and convection columns with an aspect ratio of the order of $10^5$. (The origin of the viscous estimate $\delta \sim (\mathrm{Ek})^{1/3} R_C$, which is commonplace in the literature, is explained in §5.) Finally, even in those simulations for which the forcing is modest, it has long been recognised that some of the helicity comes not from the boundary, but from the interior, originating from inhomogeneities in density (see, for example, Olson et al, 1999). It seems probable that, as the Ekman number decreases, this supplementary source of helicity progressively displaces Ekman pumping as the primary mechanism of helicity generation.

| Planet | Rotation period (days) | Core radius, $R_C$ ($10^3$km) | Dipole moment, **m** ($10^{22}$Am$^2$) | Mean axial field in the core, $\bar{B}_z$ (Gauss) | Planetary magnetic Reynolds number $R_\lambda = \frac{\Omega R_C^2}{\lambda}$ | Elsasser number $\Lambda = \frac{\sigma \bar{B}_z^2}{\rho\Omega}$ | Scaled mean axial field $\frac{\bar{B}_z/\sqrt{\rho\mu}}{\Omega R_C}$ |
|---|---|---|---|---|---|---|---|
| Mercury | 58.6 | 1.8 | 0.004 | 0.014 | $4.0\times10^{6}$ | $1.3\times10^{-4}$ | $5.6\times10^{-6}$ |
| Earth | 1 | 3.49 | 7.9 | 3.7 | $8.9\times10^{8}$ | 0.15 | $13\times10^{-6}$ |
| Jupiter | 0.413 | 55 | 150,000 | 18 | $1.3\times10^{11}$ | 3.6 | $5.2\times10^{-6}$ |
| Saturn | 0.444 | 29 | 4500 | 3.7 | $3.4\times10^{10}$ | 0.17 | $2.2\times10^{-6}$ |

Table 1. Approximate properties of those terrestrial planets and gas giants which are thought to have dynamos, along with the associated characteristic values of the Elsasser number, $\Lambda = \sigma B_z^2 / \rho\Omega$, and the scaled magnetic field, $\left(B_z/\sqrt{\rho\mu}\right)/\Omega R_C$. Here $\sigma$ is the electrical conductivity, $\lambda = 1/\mu\sigma$ is the magnetic diffusivity, and $\rho$ the density. The two dimensionless measures of magnetic field are based on the mean axial field strengths in the cores. We use the crude estimates of $\lambda \sim 1\mathrm{m}^2/\mathrm{s}$ and $\rho \sim 10^4\,\mathrm{kg/m}^3$ for the terrestrial planets, and $\lambda \sim 4\mathrm{m}^2/\mathrm{s}$ and $\rho \sim 10^3\,\mathrm{kg/m}^3$ for the gas giants (though estimates of $\lambda$ for the gas giants vary considerably).

It would seem, therefore, that if the dynamo cartoon that has emerged from the simulations is to be extended to the planets, we are obliged to identify a source of helicity in planetary cores which is robust, independent of viscosity, and internally driven (i.e. not reliant on the presence of a mantle). Above all, this source must yield a helicity distribution which is non-random and antisymmetric about the equator, ideally negative in the north and positive in the south, in accordance with the simulations. Perhaps the most obvious source of helicity are helical waves supported by the Coriolis force; either slow magnetostrophic waves or fast inertial waves. The former possibility has been considered by, say, Shimizu & Loper (2000) and Moffatt (2008), whereas here we investigate the latter option. Of the two, inertial waves have the advantage that the dynamo is able grow from a weakly magnetic state, say after a global field reversal.

In this paper we propose that inertial waves, driven by the equatorial heat flux in the outer core, are the primary source of the helical motion in the Earth (and possibly other planets), and may also act to supplement the helicity driven by Ekman pumping in the numerical simulations. In §2 we demonstrate that such waves produce the required helicity distribution outside the tangent cylinder (negative in the north and positive in the south). Moreover, we show that these waves inevitably propagate along the axis of the columnar vortices, and indeed they are the very mechanism by which the columnar vortices form in the first place and the means by which the columns subsequently evolve. Next, in §3, we calculate the emf induced by such axially propagating inertial waves and in §4 we show that, in principle, this emf is sufficient to support a self-sustaining dynamo of the $\alpha^2$ type. Finally, in §5, we derive the scaling laws for this kind of 'inertial-wave planetary dynamo'. We then compare these predictions with the imperfect simulations (imperfect in the sense that they are too viscous, rotate too slowly, and are inadequately forced), and also with what little we know about the Earth's core.

We shall see that the resulting model (perhaps we should say cartoon) is relatively robust in the sense that it is simple and invokes only familiar processes. It is also reasonably consistent with both the simulations and with our knowledge of the Earth's core. Of course, the idea that helical inertial waves are an important ingredient in dynamo action is far from new, championed in Moffatt (1970), a recurring theme in Moffatt (1978), and further explored in Olson (1981). However, we believe that this is the first time that such a self-consistent planetary dynamo has been proposed and its dynamical consequences explored in depth.

Perhaps the most surprising aspect of our dynamo cartoon is that fast inertial waves, which can transit the core on a timescale of weeks, should be an important part of a cycle in which all the other dynamical processes have timescales of hundreds of years, or longer. Consequently, the most important step in establishing the plausibility of our dynamo cartoon is to show that the continual, spontaneous generation of inertial waves in the core is not just an incidental artefact of the columnar vortices, but rather is an inescapable component of the dynamic evolution of such vortices in the outer core. To this end, as a prelude to our analysis of the dynamics of planetary cores, we first remind the reader of some classical results from the theory of rapidly-rotating fluids. The key point we seek to establish is that a slowly-evolving (quasi-geostrophic) columnar vortex in invariably immersed in a sea of inertial waves.

1.2 Classical Ideas Revisited: Transient Taylor Columns and Inertial Waves

In §2 we shall demonstrate that a quasi-geostrophic flow of the type seen in the numerical simulations of the Earth's core (that is, a flow dominated by quasi-two-dimensional columnar vortices), and which evolves slowly on a time-scale much longer than $\Omega^{-1}$, is inevitably immersed in a sea of low-frequency inertial waves. These waves propagate along the rotation axis on the fast time scale of $\Omega^{-1}$ and are the very mechanism by which the columnar vortices form and then evolve. However, it is important to note from the outset that many of these ideas are latent in the classical literature on rapidly-rotating fluids, and so it seems appropriate to first remind the reader of some of these classical concepts, and in particular of the intimate connection between Taylor columns and inertial waves. The ideas presented in this section are, of course, far from new, and so the discussion is brief and qualitative.

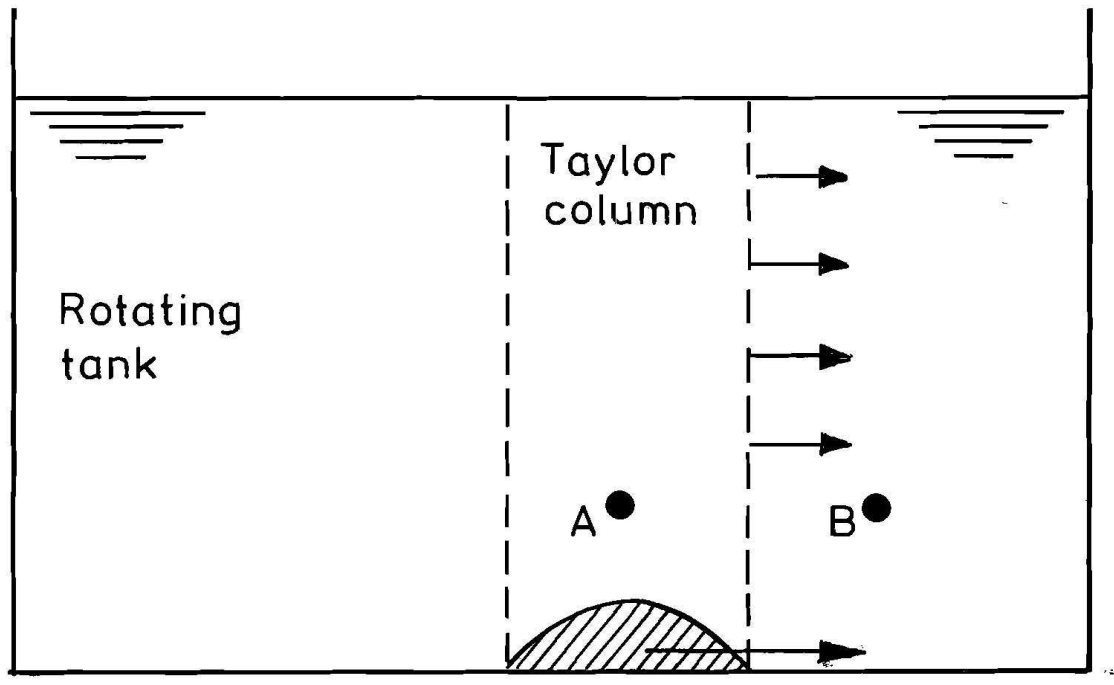

Figure 2 A small object is slowly towed across the base of a rotating tank. As the object moves it carries with it the column of fluid located between it and the upper surface of the liquid. (From Davidson, 2013a)

Let us start, then, with some classical results. Consider the experiment of Taylor, where a small object is slowly drawn across the bottom of a rapidly rotating tank which is filled with water (Figure 2). As the object moves, the column of fluid located between it and the surface of the fluid also moves, as if rigidly attached to the object. Thus, for example, a fluid particle initially at point $A$ will move across the tank, always centred above the object.

The existence of this Taylor column is usually rationalised as follows. When $\mathrm{Ro} << 1$ the curl of the Euler equation in the rotating frame of reference simplifies to the linear equation

$$\frac{\partial \boldsymbol{\omega}}{\partial t} = 2(\boldsymbol{\Omega} \cdot \nabla)\mathbf{u} , \qquad (1.2)$$

where $\mathbf{u}$ is the velocity in the rotating frame and $\boldsymbol{\omega} = \nabla \times \mathbf{u}$. If the motion is quasi-steady, then $\partial \boldsymbol{\omega} / \partial t$ may be neglected and we obtain, $(\boldsymbol{\Omega} \cdot \nabla)\mathbf{u} = 0$. Thus rapidly-rotating, quasi-steady motion is subject to the constraint that $\mathbf{u}$ is two-dimensional, in the sense that it is independent of the coordinate parallel to $\boldsymbol{\Omega}$ (the Taylor-Proudman theorem). Let us suppose $\boldsymbol{\Omega}$ points in the $z$ direction. Then (1.2) demands that $\partial u_z / \partial z = 0$ and so forbids any axial straining of fluid elements. Since a vertical column of fluid cannot be stretched or compressed, there can be no flow over the object as it drifts across the tank. Rather, the fluid must flow around the vertical cylinder which circumscribes the object, as if the Taylor column were rigid. This is the usual argument in favour of two-dimensionality. Of course, it is natural to enquire as to how the fluid lying within the Taylor column knows to move with the object, and this is where inertial waves come in.

Let us first remind ourselves of the properties of inertial waves. Application of the operator $\nabla \times (\partial / \partial t)$ to (1.2) yields the wave-like equation

$$\frac{\partial^2}{\partial t^2}\left(\nabla^2 \mathbf{u}\right) + 4(\boldsymbol{\Omega} \cdot \nabla)^2 \mathbf{u} = \mathbf{0} , \qquad (1.3)$$

which supports plane inertial waves of the form

$$\mathbf{u} = \hat{\mathbf{u}} \exp[j(\mathbf{k} \cdot \mathbf{x} - \varpi t)], \qquad \varpi = \pm 2(\mathbf{k} \cdot \boldsymbol{\Omega}) / |\mathbf{k}| ,$$

whose group velocity is

$$\mathbf{c}_g = \frac{\partial \varpi}{\partial k_i} = \pm 2 \mathbf{k} \times (\boldsymbol{\Omega} \times \mathbf{k}) / |\mathbf{k}|^3 = \pm 2 \frac{k^2 \boldsymbol{\Omega} - (\mathbf{k} \cdot \boldsymbol{\Omega})\mathbf{k}}{|\mathbf{k}|^3} . \qquad (1.4)$$

Note that low-frequency waves have $\mathbf{k} \cdot \boldsymbol{\Omega} \approx 0$ and a group velocity of $\mathbf{c}_g = \pm 2 \boldsymbol{\Omega} / |\mathbf{k}|$. Note also that, from (1.4),

$$\mathbf{c_g} \cdot \boldsymbol{\Omega} = \pm 2 k^{-3} \left[ k^2 \Omega^2 - (\mathbf{k} \cdot \boldsymbol{\Omega})^2 \right] , \qquad (1.5)$$

so that the positive sign in (1.4) corresponds to wave energy travelling upward (i.e. in the direction of $\boldsymbol{\Omega}$), while the negative sign corresponds to energy propagating downward (i.e. in the negative $z$ direction).

To make the link between Taylor columns and inertial waves it is convenient to consider an initial-value problem which is discussed in detail in Greenspan (1968) and shown schematically in Figure 3. A disc of radius $R$ is slowly moved along the axis of rotation with a speed $V$, starting at time $t$ = 0. Inevitably, low-frequency waves propagate in the $\pm \boldsymbol{\Omega}$ directions, carrying energy away from the disc at a speed $c_g \sim 2\Omega/|\mathbf{k}|$. Since the largest wavelengths travel fastest, and these have a magnitude of $|\mathbf{k}| \approx \pi / R$, we would expect to find wave-fronts located a distance $\sim (2/\pi)\Omega Rt$ above and below the disc, as shown in Figure 3. An exact solution to this problem is given in Greenspan (1968) and it turns out that the simple picture shown in Figure 3 is surprisingly accurate. At time $t$ the inertial waves generated by the disc fill a column of radius $R$ and half-length $\ell \approx (2/\pi)\Omega Rt$, and these waves carry with them the information that the disc is moving. Crucially, the fluid which lies within this column has the same axial velocity as the disc, while that lying outside the column does not know the disc is moving and so is quiescent in the rotating frame. Evidently, the inertial waves have created a form of *transient Taylor column*, whose length grows at the rate $\ell \sim c_g t$, and the role of the inertial waves is to enforce geostrophy within the column.

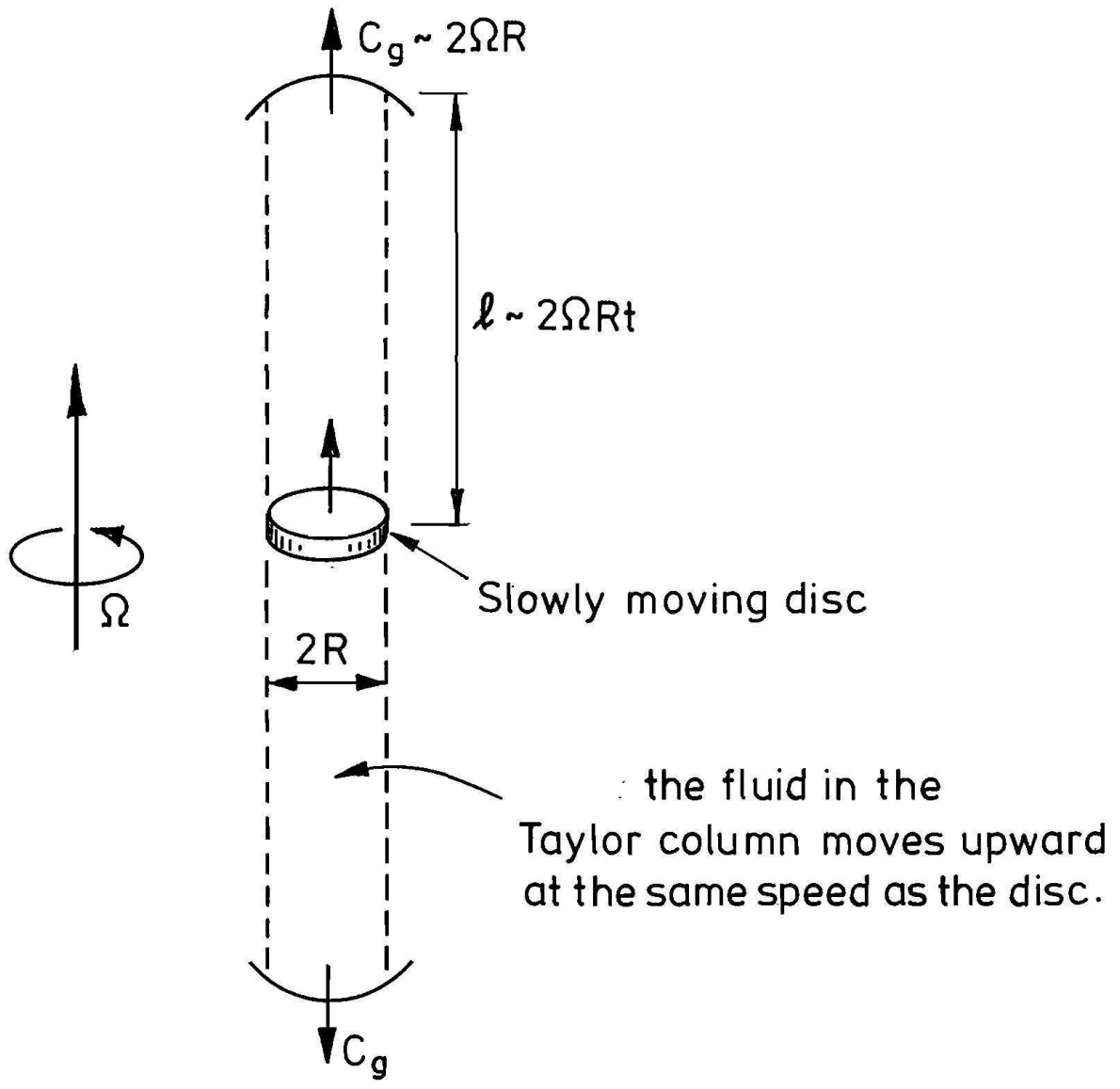

Figure 3 Formation of a transient Taylor column by inertial waves generated by a slowly moving disc. (From Davidson, 2013a)

We can now understand how the Taylor column shown in Figure 2 forms. As the object is towed slowly across the base of the tank it continually emits low-frequency inertial waves, rather like a radio antenna. These travel upward with a velocity of $\sim 2\Omega R$, and so reach the free surface on a time scale which is virtually instantaneous by comparison with the slow timescale of the movement of the object. As the inertial waves propagate upward they carry with them the information that the object is moving, and in particular the information that tells the fluid within the column to move horizontally, keeping pace with the towed object. Thus the Taylor column is continually formed and by a train of inertial waves emitted by the object. When we suppress the time derivative in (1.2) to give the Taylor-Proudman theorem, we filter out these waves. However, their long-term effect, which is the formation of the Taylor column, is still captured by the quasi-steady solution. In short, Taylor columns are formed and evolve through the continual emission of low-frequency inertial waves, and although a Taylor column may appear to evolve on a slow timescale (slow by comparison with $\Omega^{-1}$), behind this evolution we have a continual stream of fast waves whose natural time scale is $\Omega^{-1}$. In §2 we replace the towed object in Figure 2 by a buoyant blob slowly migrating across equatorial plane in the outer core.

There is one last property of inertial waves which we need to note. Inertial waves are, of course, intrinsically helical, with $\hat{\boldsymbol{\omega}} = \mp|\mathbf{k}|\hat{\mathbf{u}}$ where $\hat{\boldsymbol{\omega}}$ is the amplitude of the vorticity. It follows that the vorticity and velocity fields are parallel and in phase, with the + sign in (1.5) corresponding to negative helicity, and the – sign to positive helicity. Thus a wave packet with negative helicity will propagate upward ($\mathbf{c}_g \cdot \boldsymbol{\Omega} > 0$), while wave packets with positive helicity travel downward ($\mathbf{c}_g \cdot \boldsymbol{\Omega} < 0$). This is illustrated schematically in Figure 4 taken from Davidson (2013a), where a slab of turbulence spreads in a rotating fluid (Ro = 0.1) by emitting inertial waves. The left-hand panel is the initial condition and the right-hand one is at $\Omega t = 6$. Red represents negative helicity while green indicates positive helicity and it is clear that the upward (downward) travelling waves do indeed have negative (positive) helicity. Note that transient Taylor columns emerge spontaneously from the turbulent cloud, composed primarily of low-frequency inertial waves. This kind of spontaneous helicity generation from a localised, rotating cloud of turbulence is discussed in detail in Ranjan & Davidson, 2014.)

Returning to Figure 3, we might expect the fluid above the penny to rotate with negative $u_\theta$ as it moves upward with speed $V$ (thus ensuring $h < 0$), while the fluid below the penny might be expected to rotate with positive $u_\theta$ (giving $h > 0$), and indeed the exact solution confirms that this is so (Greenspan, 1968).

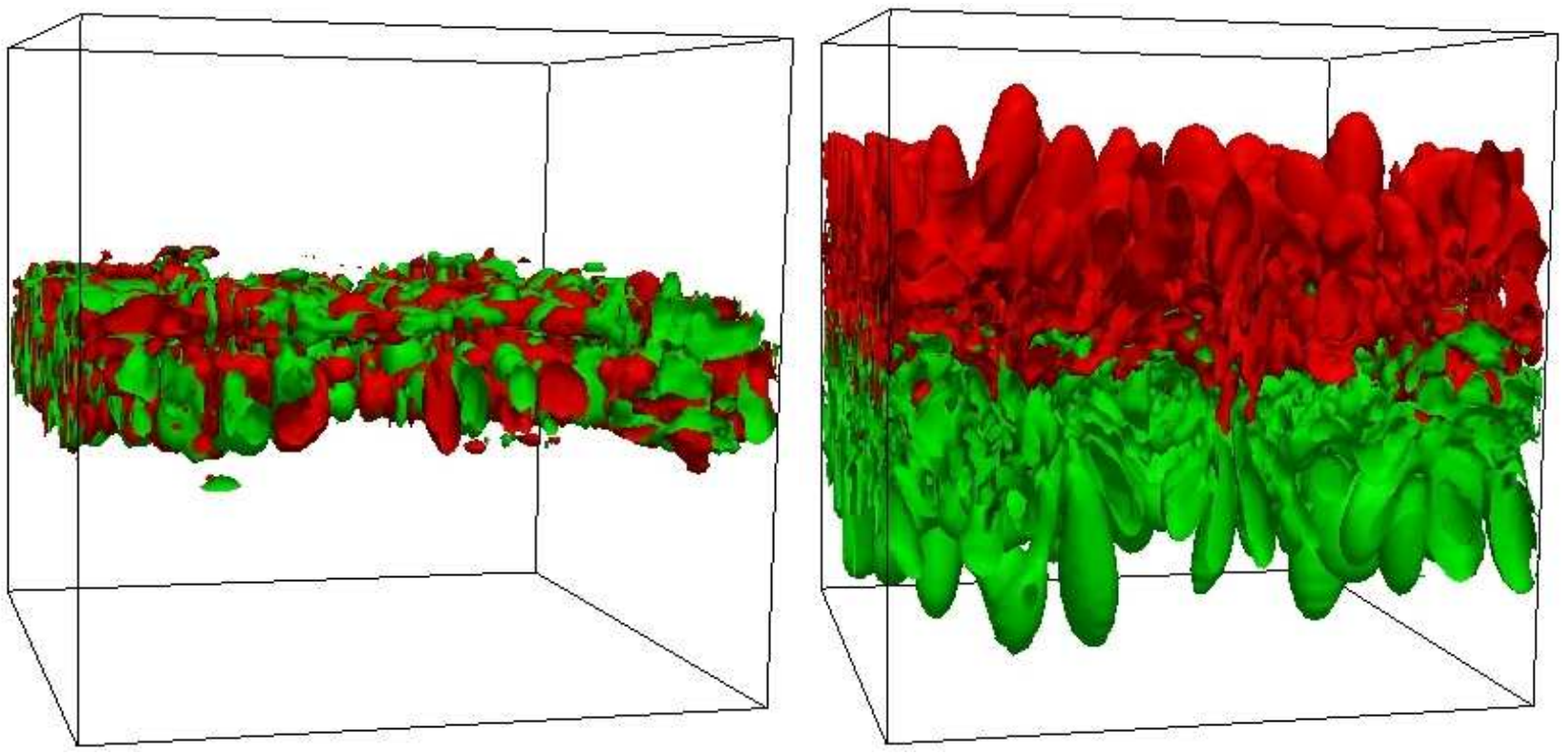

Figure 4 A slab of turbulence spreads in a rotating fluid (Ro = 0.1) by emitting inertial waves. The left-hand panel is the initial condition and the right-hand one is at $\Omega t = 6.$ Red marks negative helicity and green positive helicity. (From Davidson, 2013a.)

The picture that emerges from these simple, classical model problems is that slowly-evolving quasi-geostrophic flows are invariably immersed in a stream of low-frequency, helical inertial waves which propagate along the rotation axis, and it is the spontaneous emission of these waves which allows a quasi-geostrophic flow to evolve on the slow timescale. It is only in the extreme case of a strictly steady geostrophic flow that the inertial waves vanish.

In the context of the outer core of the Earth, quasi-geostrophic flows are established by the buoyancy field, and outside the tangent cylinder this field is most active in the equatorial plane, where there is a continual flux of buoyant material from the inner core to the mantle, possibly in the form of a radial turbulent plume. Indeed, if we were to picture the mid-plane in Figure 4 as the turbulent equatorial plane in the outer core, perhaps we have an immediate impression of how helicity might propagate in an organised manner both north and south of the equator. Of course, to substantiate such an idea we need to do a lot more work. In particular, before returning to the core of the Earth, let us consider another model problem, which is the formation of transient Taylor columns from a localised blob of buoyant fluid.

## 2. The Continual, Spontaneous Generation of Inertial Waves From Buoyant Blobs

2.1 Cyclone-Anticyclone Formation through Buoyancy-driven Inertial Waves

In most of the numerical simulations of the geodynamo it is observed that the heat flux through the outer core is concentrated in the equatorial plane (outside the tangent cylinder), and near the rotation axis (inside the tangent cylinder). Various explanations have been offered for this, and we shall provide one interpretation of this important observation shortly. In the meantime, since we are interested in a dynamo operating outside the tangent cylinder, it is natural to focus attention on the equatorial plane as the primary energy source for the dynamo. Consider, therefore, an isolated blob of buoyant material of scale $\delta$ that has left the inner core and is slowly migrating outward across the equatorial plane. We adopt local Cartesian coordinates centred on the buoyant blob, with $z$ pointing to the north, $x$ radially outward, and $y$ in the azimuthal direction, i.e. our cylindrical polar coordinates $(r,\ \theta,\ z)$ are replaced locally by $(x,\ y,\ z)$. We model the fluid as Boussinesq and ignore all secondary effects, such as a background magnetic field or stratification, and focus on the Coriolis and buoyancy forces only. In short, we replace the penny in Figure 3 by a buoyant blob. (We shall incorporate a magnetic field in the next section.) The questions we ask are: does the buoyant blob create transient Taylor columns analogous to those shown in Figure 3, and what is the associated dispersion pattern for the inertial waves?

The governing equation at low Ro is

$$\frac{\partial \mathbf{u}}{\partial t} = 2\mathbf{u} \times \mathbf{\Omega} - \nabla(p/\rho) + c\mathbf{g}, \quad c = \rho'/\rho, \tag{2.1}$$

where $\mathbf{g} = -g\hat{\mathbf{e}}_x$ is the local gravitational acceleration, $\rho'$ is the density perturbation (a negative quantity) and $\rho$ the background density. Equivalently, in terms of vorticity we have

$$\frac{\partial \boldsymbol{\omega}}{\partial t} = 2(\mathbf{\Omega} \cdot \nabla)\mathbf{u} + \nabla c \times \mathbf{g}. \tag{2.2}$$

Now $\rho'$ is governed by a simple advection diffusion equation and so evolves on a slow time scale set by the magnitude of $\mathbf{u}$. Inertial waves, on the other hand, evolve on the timescale of $\Omega^{-1}$. It follows that, at low Ro, we may treat $\rho'$ as quasi-steady as far as the initiation of inertial waves is concerned. With $\rho'$ constant, application of the operator $\nabla \times (\partial/\partial t)$ to (2.2) yields

$$\frac{\partial^2}{\partial t^2}\left(\nabla^2 \mathbf{u}\right) + (2\mathbf{\Omega} \cdot \nabla)^2 \mathbf{u} = (2\mathbf{\Omega} \cdot \nabla)(\mathbf{g} \times \nabla c), \tag{2.3}$$

and we see that the buoyancy term acts as a continual source of low-frequency inertial waves, analogous to the penny in Figure 3. These waves necessarily propagate in the $\pm\mathbf{\Omega}$ directions, carrying energy away from the buoyant blob. So, after a time $t$, we find wave-fronts located at $z \sim \pm\Omega\delta t$ above and below the blob, and within the cylindrical region defined by $z \sim \pm\Omega\delta t$ we have low-frequency inertial waves which have originated from the buoyant blob. We might expect these waves to create transients Taylor columns similar to those shown in Figure 3, and this is confirmed by (2.2) which, when the wave frequency is low, demands $(\mathbf{\Omega} \cdot \nabla)\mathbf{u} \approx 0$ outside the buoyant blob. We conclude that quasi-two-dimensional columnar vortices (i.e. transient Taylor columns) spontaneously emerge from the buoyant blob, much as they did from the impulsively started penny in Figure 3. (Some aspects of this kind of transient Taylor column formation, driven by buoyancy, are discussed in, for example, Loper, 2001, and Siso-Nadal & Davidson, 2004.)

To determine the approximate structure of these transient Taylor columns we must consider the vertical 'jump conditions' across the buoyant blob after the initial passage of inertial waves. Since $\rho'$ is quasi-steady and the inertial waves are of low frequency, (2.2) within the buoyant blob reduces to

$$2(\mathbf{\Omega}\cdot\nabla)\mathbf{u} + \nabla c \times \mathbf{g} \approx 0, \qquad (2.4)$$

or equivalently,

$$2(\mathbf{\Omega}\cdot\nabla)\boldsymbol{\omega} \approx \mathbf{g}\nabla^2 c - (\mathbf{g}\cdot\nabla)\nabla c. \qquad (2.5)$$

From (2.4) we find that the integrated vertical jump conditions across the blob are $\Delta u_x \approx 0$, $\Delta u_y \approx 0$, and

$$\Delta u_z \approx -\frac{g}{2\Omega}\int(\partial c/\partial y)dz, \qquad (2.6)$$

while (2.5) yields the vorticity jump condition

$$\Delta\omega_z \approx 0. \qquad (2.7)$$

From (2.7) we see that a cyclonic columnar vortex below the buoyant blob corresponds to a cyclonic vortex above the blob, while an anticyclonic columnar vortex below corresponds to a anticyclonic vortex above. Moreover, for a Gaussian-like buoyant blob, (2.6) tells us that $\Delta u_z$ is positive for $y < 0$ and negative for $y > 0$. It follows that $u_z$, which is antisymmetric about the plane $z = 0$, diverges from $z = 0$ for $y < 0$ (i.e. is positive for $z > 0$ and negative for $z < 0$) and converges to $z = 0$ for $y > 0$ (i.e. is negative for $z > 0$ and positive for $z < 0$). When combined with the requirement that upward propagating inertial waves have negative helicity, while downward propagating waves have positive helicity, we conclude that the inertial wave dispersion pattern consists approximately of:

(i) a pair of cyclonic and anticyclonic columnar vortices above the blob;
(ii) a matching pair of cyclonic and anticyclonic columnar vortices below the blob;
(iii) the cyclones above and below are located at the same value of $y$ (same azimuthal angle), while the anticyclones are also located at coincident azimuthal angles;
(iv) the anticyclones are located at negative $y$ (smaller azimuthal angle), and the cyclones at positive $y$ (larger azimuthal angle).

This general structure is illustrated in Figure 5, which shows a numerical simulation of the velocity field generated by a buoyant blob (of initially Gaussian profile) which slowly migrates radially outward under the influence of gravity while generating low-frequency inertial waves which propagate along the rotation axis (vertical in the image). The image shown corresponds to a time $\Omega t = 8$ after the release of the buoyant blob and the Rossby number is Ro = 0.1. The top image is coloured by helicity and the bottom one by vertical vorticity. The general structure is more or less as anticipated above, although there is a significant amount of cyclonic vorticity embedded within the anticyclone and vice versa. There are also small regions where the helicity is not of the expected sign, which may be due to the superposition of inertial waves, as discussed in Ranjan & Davidson (2014). (The simulation was performed using the pseudo-spectral code described Ranjan & Davidson, 2014, with a Courant condition based on the group velocity of inertial waves.)

It is striking that the dispersion pattern from a buoyant blob sitting in the equatorial plane is surprisingly similar to the helical flow set up by viscous Ekman pumping associated with the columnar vortices interacting with the mantle (as seen in the more weakly forced simulations). In both cases we have quasi-two-dimensional columnar vortices whose helicity is negative in the north and positive in the south, and in both cases the basic building block is a cyclonic vortex (above and below the equator) coupled to an adjacent anticyclonic vortex located at a slightly different azimuthal angle. The implication is that, in the more strongly forced simulations, the usual pattern of alternating (if irregular) cyclones and anticyclones in the outer core could be associated with random inhomogeneities in and around the equatorial plane. The main difference, of course, is that in the first case (Ekman pumping) the helicity is generated at the mantle, while in the second the helicity originates from the equatorial plane and the columnar vortices need not interact with the mantle. Indeed, we shall now show that, when the damping effect of a magnetic field is taken into consideration, the columnar vortices which originate from the equatorial plane are likely to be severely depleted by the time they reach the mantle.

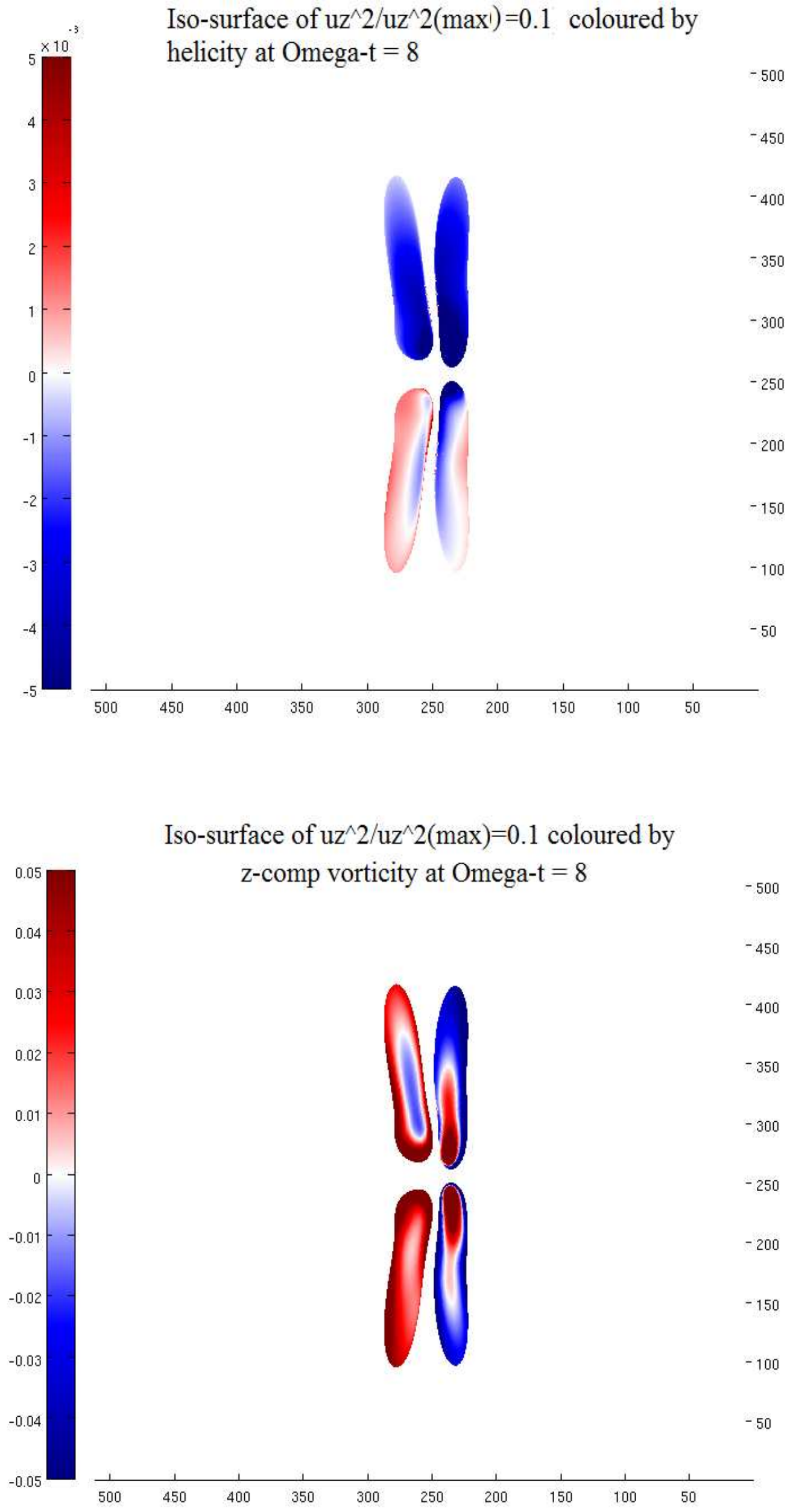

Figure 5. The structure of the dispersion pattern of inertial waves emanating from a buoyant blob which is slowly migrating across the equatorial plane towards the mantle. The top image is coloured by helicity and the bottom one by vertical vorticity.

## 2.2 Magnetic Damping of Transient Taylor Columns in the Outer Core

We close §2 by estimating the influence of magnetic damping on the transient Taylor columns generated by our buoyant blob. The first point to note it that, based on a characteristic velocity of 0.2 mm/s, and a magnetic diffusivity of $\lambda = 1\mathrm{m}^2/\mathrm{s}$, the global magnetic Reynolds for the Earth has a value of $R_m = uR_C/\lambda \sim 700$. The small scales, however, have a much smaller magnetic Reynolds number. Typical estimates of the small scales are around $\delta \sim 1\,\mathrm{km}$ - $10\,\mathrm{km}$, and we shall see that such estimates are more or less consistent with our scaling analysis. Taking $\delta = 5\,\mathrm{km}$ gives $\hat{R}_m = u\delta/\lambda = 1$ and it becomes clear that induction at the small scales is best handled using the low-$R_m$ approximation, at least for the Earth. (The low-$R_m$ approximation is reasonably good up to $R_m = 1$, as discussed in Davidson, 2013a.)

The essence of the low-$R_m$ approximation, apart from the fact that any locally induced field is much weaker than the globally imposed field, is that Ohm's law simplifies to

$$\mathbf{J} = \sigma(-\nabla\phi + \mathbf{u}\times\mathbf{B}),$$

for some potential $\phi$, from which $\nabla\times\mathbf{J} = \sigma\nabla\times(\mathbf{u}\times\mathbf{B})$. If the imposed magnetic field (which in our case we might take as the axisymmetric part of the planetary field) may be taken as locally uniform on the scale of the motion, then this simplifies to

$$\nabla\times\mathbf{J} = \sigma(\mathbf{B}\cdot\nabla)\mathbf{u}. \qquad (2.8)$$

From this we obtain a simple estimate of the Ohmic dissipation per unit mass,

$$\frac{\mathbf{J}^2}{\rho\sigma} \sim \frac{\sigma\mathbf{B}^2}{\rho}\left(\frac{\delta_{\min}}{\delta_B}\right)^2\mathbf{u}^2 \sim \left(\frac{\delta_{\min}}{\delta_B}\right)^2\frac{\mathbf{u}^2}{\tau}, \qquad (2.9)$$

where $\delta_B$ is the characteristic length-scale of the motion in the direction of the imposed magnetic field, $\delta_{\min}$ is the minimum characteristic length-scale of the motion, and $\tau = \left(\sigma\mathbf{B}^2/\rho\right)^{-1}$ is known as the Joule dissipation time.

The other important thing about the low-$R_m$ approximation is that Alfven waves become over-damped and manifest themselves as a slow diffusive growth of any localised disturbance along the imposed magnetic field lines. For example, jets or plumes diffuse along the **B**-lines according to

$$\delta_B \sim \delta_{\min}(t/\tau)^{1/2}, \qquad (2.10)$$

where $t$ is the time the disturbance has spent in the magnetic field. Thus jets and plumes rapidly adopt sheet-like structures elongated in the direction of **B**. (This kind of anisotropic, diffusive growth of localised disturbances is discussed in detail in, for example, Davidson, 2013a.) Note that substituting (2.10) into (2.9) yields

$$\frac{\mathbf{J}^2}{\rho\sigma} \sim \frac{\mathbf{u}^2}{t},$$

so that the diffusive elongation of the flow in the direction of **B** offsets to some degree the Ohmic dissipation and changes the characteristic dissipation timescale from $\tau$ to $t$.

Let us now consider how such a magnetic field might influence the growth of a transient Taylor column emerging from the equatorial plane. The low $R_m$ damping of transient Taylor columns is considered in Siso-Nadal et al, 2003, and Siso-Nadal & Davidson, 2004. In brief, the damping is crucially dependant on the relative orientation of $\mathbf{\Omega}$ and **B**. The key is (2.8), combined with the fact that axial velocity gradients within the transient Taylor column are very small, since the flow there is locally quasi-geostrophic. If $\mathbf{\Omega}$ and **B** are parallel then velocity gradients parallel to **B** are small, at least within the Taylor column, and so (2.8) then tells us that there is very little dissipation, except perhaps near the top and bottom of the growing Taylor column where velocity gradients are larger. However, if $\mathbf{\Omega}$ and **B** are mutually perpendicular, then the velocity gradients parallel to **B** (normal to the Taylor column) are large and energy is dissipated on a time scale of *t*.

Let us consider order of magnitudes, taking the buoyant blob size as $\delta = 10\,\text{km}$ and the characteristic field strength as 3.7 Gauss, which is the mean axial field strength in the core of the Earth (Table 1). The time taken for an undamped transient Taylor column to reach the mantle, starting from the equatorial plane, is the around $R_C / \Omega\delta$, which is ~36 days. By contrast, the Joule dissipation time associated with the radial and azimuthal magnetic field components is around $\tau = 25\,\text{hours}$. It follows that a growing Taylor column experiences a rapid diffusive growth of $\delta_\text{B}$ on entering a region of large radial or azimuthal magnetic field, with $\delta_\text{B}$ growing on a timescale of $\tau$ and the kinetic energy falling off at a rate proportional to *t*. By contrast, regions of the core in which the magnetic field is predominantly axial provide relatively little resistance to the formation and growth of transient Taylor columns.

The extent to which the Taylor columns are distorted in the transverse plane by the radial or azimuthal magnetic field components can be estimated as follows. The time it takes to reach the mantle from the equator is $t \sim R_C / \Omega\delta_{\min}$, in which time $\delta_\text{B}$ will grow by an amount $\delta_B \sim \delta_{\min}(t/\tau)^{1/2}$, which yields

$$\frac{\delta_B^2}{\delta_{\min} R_C} \sim \frac{1}{\Omega\tau} \sim \frac{\sigma B^2}{\rho\Omega} = \Lambda\,, \qquad (2.11)$$

where $\Lambda$ is the Elsasser number based on the radial or azimuthal magnetic field components. We shall return to (2.11) in §5.

These order of magnitude estimates have two immediate consequences. First, we can rationalise why the radial heat flux in the outer core tends to be concentrated near the equatorial plane, since the mean magnetic field there is purely axial, in that the azimuthally averaged field yields a mean $B_\theta$ which is antisymmetric about the equator (hence zero on the equator) , and a mean $B_r$ which is also zero on the equator. (Recall that we use cylindrical polar coordinates $(r,\, \theta,\, z)$.) In short, the equatorial plane is characterised by the fact that small-scale columnar convection is relatively free from magnetic damping, whereas elsewhere it is strongly damped. The second conclusion is that transient Taylor columns that originate from the equatorial plane will initially propagate relatively freely until such time that they encounter strong radial or azimuthal fields, after which they become rapidly damped and highly anisotropic in the transverse plane. The implication is that the columnar vortices will be largely dissipated by the time they reach the mantle.

Given these qualitative features of helicity generation on the equatorial plane, the task before us now is to determine whether or not this provides the basis for a self-sustaining dynamo which might be loosely labelled as $\alpha^2$ within the framework of mean-field electrodynamics. The first step is to quantify the magnitude of the currents induced by these transient Taylor columns as they propagate through the radial and azimuthal fields in the outer core.

## 3. The Alpha-effect Associated with Transient Taylor Columns

### 3.1 The Buoyant Blob Problem Revisited

Given the chaotic nature of the flow in a strongly forced dynamo, any formulation of the $\alpha$-effect must inevitably be statistical, and indeed we shall take this approach in §3.2. First, however, it is informative to return to the localised buoyant blob of §2.1 and consider a more deterministic problem. In particular, we shall calculate the currents induced by a single transient Taylor column as it grows out of the equatorial plane and into a region of significant radial or azimuthal field. (A related problem has been studied by Shimizu & Loper, 2000, though the details are very different.) While we have in mind a Taylor column growing from an isolated source of buoyancy, the results are more general. For example, it may be that a better picture of the radial heat flux across the equatorial plane is one of a turbulent radial plume which generates columnar Taylor columns in a random, if statistically organised, fashion. Much of the analysis below carries over to such a situation.

We shall adopt a classical mean-field approach to determining the emf induced the $\alpha$-effect, taking the background magnetic field to be locally uniform on the (transverse) scale of our helical Taylor column. While this is hard to justify in any formal sense, since there is unlikely to be any true separation of scales in planetary cores, it seems a natural starting point. For simplicity, we continue to model induction at the scale of the buoyant blob as being low-$R_m$, which may well limit us to certain classes of planets. We shall also simplify the problem by ignoring the inevitable diffusion of the Taylor column along the local mean field in accordance with (2.10), though this is an issue to which we shall return shortly. In short, the calculation below is purely kinematic and of a rather classical nature.

As in §2.1, we adopt local Cartesian coordinates centred on the buoyant blob, with $z$ pointing to the north, $x$ radially outward, and $y$ in the azimuthal direction. (Recall that we use cylindrical polar coordinates $(r,\ \theta,\ z)$ to describe the global geometry.) Let $\mathbf{B}$ be the local mean field, assumed uniform, $\mathbf{u}(x, y)$ be the velocity field in the Taylor column, whose weak $z$-dependence we ignore, and $\mathbf{b}$ be the small perturbation in magnetic field associated with the locally induced currents, $\mathbf{J}$. We shall also assume that the motion is helical, with $\mathbf{u} = \delta\,\boldsymbol{\omega}$, where the constant $\delta$ is negative if we are in the north and positive if we are in the south. From $\nabla \times \mathbf{J} = \sigma(\mathbf{B}\cdot\nabla)\mathbf{u}$ we see that $\mathbf{J}$ is also helical, with $\mathbf{J} = \delta\nabla\times\mathbf{J}$ and $\mathbf{b} = \mu\delta\,\mathbf{J}$. The locally induced emf is then

$$\langle \mathbf{u}\times\mathbf{b}\rangle = \mu\delta\langle \mathbf{u}\times\mathbf{J}\rangle = \mu\delta^2\langle \mathbf{u}\times\nabla\times\mathbf{J}\rangle = \frac{\delta^2}{\lambda}\langle \mathbf{u}\times(\mathbf{B}\cdot\nabla)\mathbf{u}\rangle, \qquad (3.1)$$

where the angled brackets represent a cross-sectional average across the Taylor column. The vector identity

$$\nabla(\mathbf{u}\cdot\mathbf{B}) = (\mathbf{B}\cdot\nabla)\mathbf{u} + \mathbf{B}\times\boldsymbol{\omega}$$

then allows us to rewrite $\mathbf{u}\times(\mathbf{B}\cdot\nabla)\mathbf{u}$ as

$$\mathbf{u}\times(\mathbf{B}\cdot\nabla)\mathbf{u} = \mathbf{u}\times\nabla(\mathbf{u}\cdot\mathbf{B}) - \mathbf{u}\times(\mathbf{B}\times\boldsymbol{\omega}) = (\mathbf{u}\cdot\mathbf{B})\boldsymbol{\omega} - \nabla\times((\mathbf{u}\cdot\mathbf{B})\mathbf{u}) - \mathbf{u}\times(\mathbf{B}\times\boldsymbol{\omega}). \qquad (3.2)$$

On substituting into (3.1), noting that $\mathbf{u}$ vanishes outside the Taylor column, and that $\mathbf{u} = \delta\boldsymbol{\omega}$, we find

$$\langle \mathbf{u}\times\mathbf{b}\rangle = \frac{\delta}{\lambda}\langle(\mathbf{u}\cdot\mathbf{B})\mathbf{u} - \mathbf{u}\times(\mathbf{B}\times\mathbf{u})\rangle = -\frac{\delta}{\lambda}\langle(\mathbf{u}^2)\mathbf{B} - 2(\mathbf{u}\cdot\mathbf{B})\mathbf{u}\rangle. \qquad (3.3)$$

Next we note that $\mathbf{b}=0$ when $\mathbf{B}$ is purely axial, and so (3.3) demands

$$\langle u_z^2\rangle = \langle u_x^2 + u_y^2\rangle, \quad \text{and} \quad \langle u_x u_z\rangle = \langle u_y u_z\rangle = 0, \tag{3.4}$$

a result which, in fact, follow directly from the assumed two-dimensional, helical structure of $\mathbf{u}$, $\mathbf{u}(x,y)=\delta\boldsymbol{\omega}$. It follows that, when there is a radial (i.e. $x$) component of the mean field, the induced emf is of the form

$$\langle \mathbf{u}\times\mathbf{b}\rangle = -\frac{\delta B_x}{\lambda}\langle 2u_y^2, -2u_x u_y, 0\rangle, \tag{3.5}$$

and when there is an azimuthal (i.e. $y$) component of the mean field, we have

$$\langle \mathbf{u}\times\mathbf{b}\rangle = -\frac{\delta B_y}{\lambda}\langle -2u_x u_y, 2u_x^2, 0\rangle. \tag{3.6}$$

Of course, we have $\langle \mathbf{u}\times\mathbf{b}\rangle = 0$ when both the radial and azimuthal components of the mean field are zero. If the buoyant blob, and hence the resulting Taylor column, happens to be symmetric about $x$ (the radial coordinate), then $\langle u_x u_y\rangle = 0$ and the induced emf simplifies to

$$\langle \mathbf{u}\times\mathbf{b}\rangle = -\frac{2\delta}{\lambda}\langle u_y^2 B_x, u_x^2 B_y, 0\rangle. \tag{3.7}$$

Reverting to global polar coordinates, we conclude that, when averaged across the cross-section of the Taylor column, the local emf is

$$\langle \mathbf{u}\times\mathbf{b}\rangle \approx -\frac{\delta\langle \mathbf{u}^2\rangle}{2\lambda}\mathbf{B}_\perp, \quad \mathbf{B}_\perp = \mathbf{B} - B_z\hat{\mathbf{e}}_z, \tag{3.8}$$

where we have assumed that $\langle u_x^2\rangle \approx \langle u_y^2\rangle$. Thus, in accordance with classical mean-field electrodynamics, we have a local mean emf given by

$$\langle \mathbf{u}\times\mathbf{b}\rangle \approx -\frac{\delta\langle \mathbf{u}^2\rangle}{2\lambda}\mathbf{B}_\perp = -\frac{\delta^2\langle h\rangle}{2\lambda}\mathbf{B}_\perp. \tag{3.9}$$

We shall see in §5 that, in principle, this is sufficiently large to support a self-sustaining dynamo.

### 3.2 The Low-$R_m$ $\alpha$-Effect for a Random Sea of Transient Taylor Columns

So far we have considered a single Taylor column and then averaged across that column. To make progress we would now have to make some assumption about the statistical distribution of such columns outside the tangent cylinder. In many ways it makes more sense to adopt a statistical approach from the outset, which we now do. It is shown in Davidson (2001) that a statistically homogeneous (but not necessarily isotropic) field of turbulence evolving in a locally uniform magnetic field induces a low-$R_m$ emf of

$$\langle \mathbf{u}\times\mathbf{b}\rangle = -\frac{1}{\lambda}\langle (\mathbf{a}\cdot\mathbf{u})\mathbf{B} - 2(\mathbf{a}\cdot\mathbf{B})\mathbf{u}\rangle, \tag{3.10}$$

where **a** is the solenoidal vector potential for **u** and the angled brackets now represent a local volume average. Assuming that **u** is everywhere helical, $\mathbf{a} = \delta \mathbf{u}$, this simplifies to

$$\langle \mathbf{u} \times \mathbf{b} \rangle = -\frac{\delta}{\lambda} \langle (\mathbf{u}^2)\mathbf{B} - 2(\mathbf{u} \cdot \mathbf{B})\mathbf{u} \rangle, \qquad (3.11)$$

which brings us back to (3.3), but with a different interpretation of $\langle \sim \rangle$. We now note that (3.4),

$$\langle u_z^2 \rangle = \langle u_x^2 + u_y^2 \rangle, \ \langle u_x u_z \rangle = \langle u_y u_z \rangle = 0,$$

follow directly from the assumed two-dimensional, helical structure of **u**, and invoke two-dimensional isotropy, which demands $\langle u_x^2 \rangle = \langle u_y^2 \rangle$ and $\langle u_x u_y \rangle = 0$. Expression (3.11) then simplifies to

$$\langle \mathbf{u} \times \mathbf{b} \rangle = -\frac{\delta \langle \mathbf{u}^2 \rangle}{2\lambda} \mathbf{B}_\perp = -\frac{\delta^2 \langle h \rangle}{2\lambda} \mathbf{B}_\perp, \qquad (3.12)$$

which is, of course, (3.9) by a different route. The advantage of expression (3.12) over (3.9) is that it can be interpreted in terms of a local volume average in the outer core, and rests simply on the statistical assumptions of: (i) local homogeneity; (ii) maximum helicity; (iii) a weak *z*-dependence of **u**; and (iv) two-dimensional isotropy. This frees us from the deterministic cartoon of individual helical Taylor columns generated by local patches of buoyancy. On the other hand, if we believe that discrete transient Taylor columns growing out of the equatorial plane are indeed the basic building block, then perhaps the more deterministic estimate (3.9) is to be preferred. Either way we have

$$\langle \mathbf{u} \times \mathbf{b} \rangle \approx \pm \frac{|\delta| \langle \mathbf{u}^2 \rangle}{2\lambda} \mathbf{B}_\perp, \qquad (3.13)$$

where the upper sign corresponds to the north and the lower to the south.

3.3 The Influence of Anisotropy Arising From Diffusion Along The Magnetic Field Lines

Before leaving the subject of the induced emf, it is instructive to consider the role of anistopy in the plane normal to *z*. The point is this. As a Taylor column grows into a region of significant radial or azimuthal magnetic field it will start to diffuse along the transverse field component, and so the cross-section of the Taylor column becomes progressively distorted into a sheet-like structure of thickness $\delta_{\min}$ (which is set by the size of the buoyant blob) and width $\delta_B$, where $\delta_B$ is measured in the direction of $B_r$ or $B_\theta$, as appropriate. Indeed, according to (2.10), $\delta_B$ grows diffusively according to $\delta_B \sim \delta_{\min}(t/\tau)^{1/2}$, as discussed in Davidson (2013a) and Siso-Nadal & Davidson (2004). So the cross-section of the Taylor column will be increasingly characterised by two transverse length-scales and two corresponding transverse velocities, $u_{\min}$ and $u_B$, with $u_{\min}/\delta_{\min} \sim u_B/\delta_B$ and $u_B \sim u_z$. The question then arises as to which of these two transverse lengths and velocities should appear in an appropriately modified version of (3.13). It is readily confirmed that the answer is

$$\langle \mathbf{u}\times\mathbf{b}\rangle \sim \pm\frac{\delta_{\min} u_{\min}^2}{\lambda}\mathbf{B}_\perp \ , \qquad (3.14)$$

where, as in (3.13), the upper sign corresponds to the north and the lower to the south.

This may be established as follows. To focus thoughts, suppose that the axial flow in the column is positive (i.e. upward) and symmetric about both axes of the column's cross-section. Suppose also that the streamlines in the transverse plane take the form of nested, flattened ellipses, with the long axis aligned with $\mathbf{B}_\perp$. We shall see that **J** and **b** have components which are either symmetric or antisymmetric about the minor axis of the elliptical cross-section, so we shall find it convenient to distinguish between the two sides of the minor axis by referring to the front and rear of the column's cross-section, where the direction of $\mathbf{B}_\perp$ takes us from the rear to the front.

Consider first the axial component of motion interacting with $\mathbf{B}_\perp$. The induced currents are determined by $\nabla\times\mathbf{J} = \sigma(\mathbf{B}\cdot\nabla)\mathbf{u}$, which yields a dipolar distribution of current which recirculates in the transverse plane and is mirror symmetric about the minor axis. The magnitude of these currents is of the order of $J_{\min} \sim \sigma B u_z(\delta_{\min}/\delta_B)^2$ and $J_B \sim \sigma B u_z(\delta_{\min}/\delta_B)$, where $J_B$ is the component of **J** parallel to $\mathbf{B}_\perp$ and $J_{\min}$ is normal to $\mathbf{B}_\perp$. From Ampere's law, these currents then induce a vertical magnetic field which is antisymmetric about the minor axis, being positive at the rear and negative at the front. Moreover, the magnitude of this vertical field is $b_z \sim (B u_z/\lambda)(\delta_{\min}^2/\delta_\mathbf{B})$, or equivalently $b_z \sim B u_{\min}\delta_{\min}/\lambda$. Consider next the interaction of the horizontal motion with $\mathbf{B}_\perp$, which is also governed by $\nabla\times\mathbf{J} = \sigma(\mathbf{B}\cdot\nabla)\mathbf{u}$. The induced current is now vertical, of magnitude $J_z \sim \sigma B u_{\min}$, and like $b_z$ it is antisymmetric about the minor axis, the signs of $J_z$ either side of the minor axis being dependent on the helicity in the Taylor column. From Amperes law, the associated magnetic field sits in the transverse plane and is mirror symmetric about the minor axis. The component of this field normal to $\mathbf{B}_\perp$ is of the order of $b_{\min} \sim (B u_{\min}/\lambda)(\delta_{\min}^2/\delta_\mathbf{B})$. From these various estimates of **b** we can calculate the order of magnitude of $\langle \mathbf{u}\times\mathbf{b}\rangle$, and after a little effort this brings us back to (3.14). Estimate (3.14) will play an important role in our scaling analysis in §5.

## 4. Cartoons For Planetary Dynamos Driven by Inertial Waves

### 4.1 An $\alpha^2$ Model Based on Buoyant Equatorial Blobs or Plumes

Let us now gather together the various threads from the preceding sections and see if we can construct an $\alpha^2$-like dynamo cartoon of the type shown in Figure 1, based on the idea of transient Taylor columns growing out of the equatorial plane, carrying their helicity with them. The key input to such a cartoon is (3.14), in the form

$$\langle \mathbf{u}\times\mathbf{b}\rangle \sim \pm\frac{\delta_{\min} u_{\min}^2}{\lambda}\mathbf{B}_\perp . \qquad (4.1)$$

Noting that $u_{\min}/\delta_{\min} \sim u_B/\delta_B$, we may rewrite this as

$$\langle \mathbf{u}\times\mathbf{b}\rangle \sim \pm\frac{\delta_{\min} u^2}{\lambda}\left(\frac{\delta_{\min}}{\delta_B}\right)^2\mathbf{B}_\perp , \qquad (4.2)$$

where from now on we shall drop the subscript $B$ on $u_B$, on the grounds that $u_B \sim u_z$ is the typical velocity in the core. The smaller velocity, $u_{\min}$, is then an auxiliary quantity, to be determined from $u_{\min} \sim u\delta_{\min}/\delta_B$.

From (4.1) or (4.2) it is clear that, in principle, we can construct an $\alpha^2$ dynamo of the form shown in Figure 1. As usual, we use cylindrical polar coordinates $(r,\ \theta,\ z)$ to describe the global geometry. Let us assume the dipole points to the north. Then the interaction of the helical columns with the mean radial field generates an emf of the form

$$\langle \mathbf{u}\times\mathbf{b}\rangle_r \sim \pm\frac{\delta_{\min} u^2}{\lambda}\left(\frac{\delta_{\min}}{\delta_B}\right)^2 B_r \sim \frac{\delta_{\min} u^2}{\lambda}\left(\frac{\delta_{\min}}{\delta_B}\right)^2 |B_r|, \tag{4.3}$$

which is zero on the equator and grows in magnitude as we move towards the mantle. The resulting poloidal current, $\mathbf{J}_p$, will be as shown in Figure 1, with a quadrupole structure which is mirror symmetric about the equator and has positive radial current at large latitudes. The magnitude of $\mathbf{J}_p$ will be of order $\sigma\langle \mathbf{u}\times\mathbf{b}\rangle_r$, and from Ampere's law this current will induce an azimuthal magnetic field of magnitude

$$|B_\theta| \sim \mu R_C |J_p| \sim \frac{R_C}{\lambda}\langle \mathbf{u}\times\mathbf{b}\rangle_r \sim \frac{R_C}{\lambda}\frac{\delta_{\min} u^2}{\lambda}\left(\frac{\delta_{\min}}{\delta_B}\right)^2 |B_r|, \tag{4.4}$$

which is antisymmetric about the equator and positive in the north. The interaction of the helical columns with this azimuthal field then generates an azimuthal emf of magnitude

$$\langle \mathbf{u}\times\mathbf{b}\rangle_\theta \sim \pm\frac{\delta_{\min} u^2}{\lambda}\left(\frac{\delta_{\min}}{\delta_B}\right)^2 B_\theta \sim \frac{\delta_{\min} u^2}{\lambda}\left(\frac{\delta_{\min}}{\delta_B}\right)^2 |B_\theta|, \tag{4.5}$$

which is positive in both the north and south. The resulting azimuthal current has magnitude $J_\theta \sim \sigma\langle \mathbf{u}\times\mathbf{b}\rangle_\theta$. Finally, from Ampere's law, $\mathbf{J}_\theta$ will support the a north-pointing dipole magnetic field of magnitude

$$|B_p| \sim \mu R_C |J_\theta| \sim \frac{R_C}{\lambda}\langle \mathbf{u}\times\mathbf{b}\rangle_\theta \sim \frac{R_C}{\lambda}\frac{\delta_{\min} u^2}{\lambda}\left(\frac{\delta_{\min}}{\delta_B}\right)^2 |B_\theta|. \tag{4.6}$$

With (4.6) we have completed the dynamo cycle. Moreover, a comparison of (4.4) and (4.6) shows that the cycle is self-sustaining provided that

$$\frac{R_C}{\lambda}\frac{\delta_{\min} u^2}{\lambda}\left(\frac{\delta_{\min}}{\delta_B}\right)^2 \sim 1,$$

or equivalently

$$\frac{R_C u}{\lambda}\cdot\frac{\delta_{\min} u}{\lambda} \sim \left(\frac{\delta_B}{\delta_{\min}}\right)^2. \tag{4.7}$$

We shall explore the consequences of this relationship in §5. However, we note immediately that, since $R_m = uR_C/\lambda >> 1$, the dynamo can be self-sustaining if either $\hat{R}_m = u\delta_{\min}/\lambda << 1$, or else $\hat{R}_m \sim 1$ and the flow in the transverse plane is highly anisotropic in the sense that $\delta_{\mathbf{B}} >> \delta_{\min}$. We shall see that, for the Earth, the latter option is more likely.

### 4.2 Other Dynamo Cartoons Based on Transient Taylor Columns

The attraction of the cartoon outlined above is that it is consistent with the observation that the heat flux outside the tangent cylinder is concentrated on the equatorial plane, and that dynamo action seems to be located outside the tangent cylinder, at least in the numerical simulations. However, before leaving the subject of mean-field dynamos whose helicity is supplied by transient Taylor columns, perhaps it is worth noting that there are other possibilities. This seems prudent, since we do not know how the results of the simulations will change as the level of forcing increases, the viscosity decreases, and the Rossby number falls. The first point to note is that the dynamo cartoon shown in Figure 1 still works if the sign of the helicity is reversed in both the north and south. The only consequence of reversing the signs of $h$ is that $B_\theta$ changes sign. So, if the Taylor columns were generated near the mantle, rather than at the equator, we could still get a dynamo. In short, the only kinematic requirement for an $\alpha^2$ dynamo operating outside the tangent cylinder is that the heat flux adopts a statistically steady pattern which is strongly non-uniform in the outer core.

The relationship between of $h$ and $B_\theta$ required for dynamo action is most easily seen from the general expression (Davidson, 2013a)

$$\frac{d}{dt}\int_{V_C}\left(R_C^2 - \mathbf{x}^2\right)B_z dV = 2\int_{V_C} r\langle \mathbf{u}\times\mathbf{b}\rangle_\theta dV - 6\lambda\int_{V_C} B_z dV\,, \tag{4.8}$$

where $B_z$ is the axisymmetric part of the axial field. When combined with the Taylor column estimate (3.12),

$$\langle \mathbf{u}\times\mathbf{b}\rangle = -\frac{\delta^2\langle h\rangle}{2\lambda}\mathbf{B}_\perp\,,$$

we find

$$\frac{d}{dt}\int_{V_C}\left(R_C^2 - \mathbf{x}^2\right)B_z dV = -\frac{\delta^2}{\lambda}\int_{V_C} r\langle h\rangle B_\theta dV - 6\lambda\int_{V_C} B_z dV\,. \tag{4.9}$$

Evidently $h$ and $B_\theta$ must have opposite signs in order to maintain a dynamo.

Finally we note that transient Taylor columns could, in principle, also yield the helicity needed to sustain an $\alpha-\Omega$ dynamo. The evidence of the numerical simulations suggests that, if such a dynamo were to exist, it would probably reside inside the tangent cylinder where strong buoyant upwellings can drive a significant $\Omega$ effect. Indeed, a cartoon of just such an $\alpha-\Omega$ dynamo operating within the tangent cylinder is sketched in Davidson (2004), though there is no evidence of this kind of dynamo in the simulations.

## 5. Scaling Laws for an Inertial Wave Dynamo

### 5.1 Theoretical Scaling Laws

We now turn to the scaling laws which accompany an $\alpha^2$ dynamo of the type discussed in §4.1. To this end we must examine the force balance within the transient Taylor columns and combine this with the kinematic requirement (4.7). We shall make two key assumptions in what follows. First, we shall continue to make the assumption that induction at the scale of the transient Taylor columns can be modelled as a low-$R_m$ process. This automatically sets the dissipation scale in the outer core to the transverse scale of the Taylor columns, and ensures that $\lambda$ is an important scaling parameter. This may be contrasted with other proposed scaling laws in which $\lambda$ is assumed to be unimportant and the dissipation scale is taken to be much smaller than the transverse scale of the columnar structures (see, for example, Davidson, 2013b). The essential distinction between the two classes of scaling laws is that in the latter case induction within the columnar vortices is assumed to operate at moderate to large $R_m$. Our second assumption is that most of the transient Taylor columns straddle the bulk of the outer core, so that their length scales on $R_C$. Admittedly, both of these assumptions are difficult to validate. For example, the assumption of low-$R_m$ induction within the columnar vortices may seem plausible for the Earth, as discussed in §2.2, but it is far from clear that it is appropriate for the gas giants. Perhaps the most we can do in this regard is look for self-consistency in the model and check that the predicted values of $R_m$ are order one or less.

At this point it is convenient to replace the scaled density perturbation, $c = \rho' / \rho$, by the equivalent temperature perturbation, $-\beta T'$, and introduce the time-averaged rate of production of energy per unit mass, $\mathrm{P} = -\beta \overline{T'\mathbf{u}} \cdot \mathbf{g}$, which includes contributions from both the steady-on-average convection and the turbulence. (The over-bar here represents a time average and $\beta$ is the thermal expansion coefficient.) Noting that the time-averaged convective heat flux per unit area, $\mathbf{q}_T$, is given by $\mathbf{q}_T / \rho c_p = \overline{T'\mathbf{u}}$, we conclude that

$$\mathrm{P} = \frac{g\beta}{\rho c_p} q_T , \tag{5.1}$$

where $q_T$ is the radial convective heat flux out through the core and $g = |\mathbf{g}|$. Since the convective heat flux varies throughout the core, it is convenient to introduce $\overline{\mathrm{P}}$, the volume average of P over the outer core, which is related to the net convective heat flux out of the core, $Q_T$, by

$$\overline{\mathrm{P}} \sim \frac{g\beta}{\rho c_p} \frac{Q_T}{4\pi R_C^2} . \tag{5.2}$$

We may think of $\overline{\mathrm{P}}$, $R_C$, $\Omega$ and $\lambda$ as given parameters, and $B$, $u$, $\delta$ and $\delta_B$ as the dependant variables to be determined by the scaling analysis.

In terms of dimensionless groups, we may regard

$$\Pi_P = \frac{\overline{\mathrm{P}}}{\Omega^3 R_C^2} , \qquad R_\lambda = \frac{\Omega R_C^2}{\lambda} , \tag{5.3}$$

as the independent groups, and

$$\Lambda = \frac{\sigma B^2}{\rho\Omega} , \quad \mathrm{Ro} = \frac{u}{\Omega R_C} , \frac{\delta_{\min}}{R_C} , \frac{\delta_B}{R_C} , \tag{5.4}$$

as dimensionless measures of $B$, $u$, $\delta_{\min}$ and $\delta_B$. In fact, recently it has become conventional to adopt

$$\Pi_B = \frac{B/\sqrt{\rho\mu}}{\Omega R_C} , \tag{5.5}$$

rather than $\Lambda$, as the dimensionless measure of *B*, which is clearly related to the Elsasser number through $\Lambda = \Pi_B^2 R_\lambda$. From these four dependant dimensionless groups me may construct other useful dimensionless quantities, such as $R_m = uR_C/\lambda$ and $\hat{R}_m = u\delta_{\min}/\lambda$.

Let us now consider the governing equations at our disposal. Balancing the curl of the Coriolis force, $2(\mathbf{\Omega}\cdot\nabla)\mathbf{u}$, against the curl of the buoyancy forces we have

$$\frac{\overline{\mathrm{P}}}{u\delta_{\min}} \sim \frac{\Omega u}{R_C}\ , \tag{5.6}$$

or, in dimensionless form,

$$\mathrm{Ro} \sim \Pi_p^{1/2}\sqrt{R_C/\delta_{\min}}\ . \tag{5.7}$$

To this we must add (2.11),

$$\frac{\delta_B^2}{\delta_{\min} R_C} \sim \frac{\sigma B^2}{\rho\Omega} = \Lambda\ , \tag{5.8}$$

which can now be reinterpreted as a balance between the curl of the Coriolis force, $2(\mathbf{\Omega}\cdot\nabla)\mathbf{u}$, and the curl of the low-$R_m$ Lorentz force. Finally we have the requirement that the dynamo is self-sustaining, which demands

$$\frac{R_C\,u}{\lambda}\cdot\frac{\delta_{\min}\,u}{\lambda} \sim \left(\frac{\delta_B}{\delta_{\min}}\right)^2 . \tag{5.9}$$

Expressions (5.7)-(5.9) are the basis of our scaling laws. Since we have four unknowns and only three equations it is immediately apparent that the system is not closed and that consequently we have overlooked some important physical process. Arguably, the missing information is the procedure by which $\delta_{\min}$ is set. There are a number of ways forward at this point.

Some authors (e.g. King & Buffett, 2013) advocate a viscous force balance within the columnar convection cells in which $\nu\nabla^2\boldsymbol{\omega}$ is of the order of $2(\mathbf{\Omega}\cdot\nabla)\mathbf{u}$. This then fixes the column width as

$$\delta_{\min}/R_C \sim \left(\nu/\Omega R_C^2\right)^{1/3} \sim (\mathrm{Ek})^{1/3}. \tag{5.10}$$

If we adopt this, along with the other two force balances, then the system is closed. However, while this may well be valid for the more viscous numerical simulations, it seems improbable for the interior of a planet, as discussed in §1.

An alternative approach is to accept that $\delta_{\min}$ is fixed by some process which we have not modelled, such as the formation of the equatorial radial plume at the inner core boundary, and simply accept that our system of equations is under specified. To plug the gap we might take the measured value of $\Lambda$, or equivalently $\Pi_B$, as an empirical input on the grounds that $\Lambda$ is reasonably well known for the planets (Table 1). With given values of $\Pi_P$, $R_\lambda$ and $\Pi_B$, we can then calculate all the remaining parameters and check that the results are self-consistent (e.g. $\hat{R}_m = u\delta_{\min}/\lambda \le 1$) and in line with what we know about the Earth and with the (imperfect) numerical simulations.

With this strategy in mind, we can manipulate our three governing equations into a more convenient form. First, eliminating $\delta_B$ from (5.8) and (5.9) yields

$$\hat{R}_m = \frac{u\delta_{\min}}{\lambda} \sim \Lambda^{1/2}. \qquad (5.11)$$

Second, eliminating $\delta_{\min}$ from (5.7) and (5.11) we find

$$\text{Ro} \sim \frac{\Pi_P R_\lambda}{\Lambda^{1/2}} \sim \frac{\Pi_P R_\lambda^{1/2}}{\Pi_B}, \qquad (5.12)$$

which combined with (5.11) gives,

$$\frac{\delta_{\min}}{R_C} \sim \frac{\Pi_B^2}{\Pi_P R_\lambda}. \qquad (5.13)$$

Third, (5.8) can be combined with (5.13) to yield

$$\frac{\delta_B}{\delta_{\min}} \sim \Pi_P^{1/2} R_\lambda. \qquad (5.14)$$

Equations (5.12) – (5.14) between them specify $u$, $\delta_{\min}$ and $\delta_B$ in terms of $\Pi_P$, $R_\lambda$ and $\Pi_B$, as required.

Note that (5.13) can be rearranged as

$$\overline{\text{P}} \sim \frac{B^2}{\rho\mu}\frac{\lambda}{\delta_{\min} R_C}, \qquad (5.15)$$

which represents the energy balance $\overline{\text{P}} = J^2 / \rho\sigma$. This tells us that the magnetic energy is independent of $\Omega$ (as advocated by Christensen et al, 2009), provided that $\delta_{\min}$ is independent of the rotation rate. A similar conclusion was reached in Davidson (2013b).

Perhaps it is worth summarizing the regime in which our model operates. We require: (i) $u\delta_{\min}/\lambda < 1$ (i.e. $\Lambda^{1/2} < O(1)$) so that induction in the columns can be modelled as a low-$R_m$ process; (ii) $u/\Omega\delta_{\min} < 1$, or equivalently $\text{Ro}^3/\Pi_P < O(1)$, so that inertial waves can propagate on the scale of $\delta_{\min}$; and (iii) $P_m < 1$ so that Joule dissipation dominates over viscous dissipation. All three of these conditions are likely to be met in the core of the Earth, but do not hold in many of the simulations.

Finally, we note that, in cases where $\delta_{\min}$ is set by the viscous scale $\delta_{\min}/R_C \sim (\text{Ek})^{1/3}$, but helicity generation is dominated by inertial waves, which may be the case in some of the numerical simulations (though presumably not in planets), (5.7) and (5.13) become

$$\text{Ro} \sim \Pi_p^{1/2}\text{Ek}^{-1/6} \sim \Pi_p^{4/9} P_m^{-1/6}\left(\frac{\overline{\text{P}}^{1/3} R_C^{4/3}}{\lambda}\right)^{1/6}, \qquad (5.16)$$

$$\Pi_B \sim \Pi_p^{1/2}\text{Ek}^{-1/3} P_m^{1/2} \sim \Pi_p^{7/18} P_m^{1/6}\left(\frac{\overline{\text{P}}^{1/3} R_C^{4/3}}{\lambda}\right)^{1/3}, \qquad (5.17)$$

where $P_m$ is the magnetic Prandtl number. The regime in which this holds can be estimated from combining (5.16) with $\delta_{\min}/R_C \sim (\text{Ek})^{1/3}$ and the need for $u/\Omega\delta_{\min} < 1$. This yields $\Pi_p/\text{Ek} < O(1)$, or equivalently, $\overline{\text{P}}/\Omega^2\nu < O(1)$. At yet lower rotation rates the helicity is unlikely to be generated by inertial waves and we must fall back on, say, Ekman pumping. At yet higher values of $\overline{\mathrm{P}}/\Omega^2\nu$ (higher forcing, weaker rotation) we might envisage that neither inertial waves nor Ekman pumping are effective sources of helicity, in which case the dipolar dynamo will fail, and the likely outcome will be a so-called small scale dynamo, in which magnetic energy is generated at the small scales only. This is reminiscent of the findings of Christensen & Aubert (2006).

## 5.2 A Comparison with the Known Properties of the Planets and the Numerical Simulations

Let us now compare these predictions with the known properties of the Earth. The first problem here is to estimate a representative field strength in the core. Although the spatially averaged axial field is $\overline{B}_z$= 3.7 Gauss (Table 1), most people believe that the rms field strength is an order of magnitude higher. This is based partly on the observed frequency of torsional oscillations in the core, in which the field acts as a magnetic spring. So let us adopt the estimate $B$ = 30 Gauss. Next, in line with earlier studies, we shall take $\Pi_P = 5\times10^{-14}$ ($Ra_Q \approx 2\times10^{-13}$), corresponding to a convective heat flux at the core-mantle boundary of around 2 T Watts. Finally, we assume $R_\lambda = 1.26\times10^9$, based on the recent estimate of $\lambda = 0.7\,\mathrm{m^2/s}$ for the Earth's core.

By necessity, we take all pre-factors as unity in our scaling laws, so that numerical estimates of particular quantities must be regarded as indicative of an order of magnitude only. The tentative values of $B$, $\Pi_P$ and $R_\lambda$ above yield the following estimates:

$$\mathrm{Ro} \sim 1.5\times10^{-5} \text{ (i.e. } u \sim 4\,\mathrm{mm/s}), \qquad u/\Omega\delta_{\min} \sim 0.08,$$
$$\delta_{\min}/R_C \sim 1.9\times10^{-4} \text{ (i.e. } \delta_{\min} \sim 0.7\,\mathrm{km}), \qquad \delta_B/\delta_{\min} \sim 260,$$
$$\hat{R}_m = u\delta_{\min}/\lambda \sim 3.6, \qquad R_m = uR_C/\lambda \sim 19\times10^3.$$

The predicted Rossby number is a factor of ~10 larger than most values quoted in the literature, and so our scaling laws have overestimated $u$ by a corresponding factor. (Actually, we shall see shortly that the pre-factor in (5.12) is likely to be small and this may, in part, offset the apparent overestimate of $u$.) It is reassuring that $u/\Omega\delta_{\min} < 1$, since we require that inertial waves can propagate freely at the small scales, which is possible only if $u/\Omega\delta_{\min} < 1$. The estimate $\delta_{\min} \sim 1\mathrm{km}$ seems not unreasonable, though $\hat{R}_m = u\delta_{\min}/\lambda \sim 3.6$ is a little higher then we would have wanted in order to justify the low-$R_m$ approximation. In fact, both $\hat{R}_m$ and $R_m$ are significantly larger than their usual estimates, which is almost certainly because we have overestimated $u$. Finally, $\delta_B/\delta_{\min} \sim 260$ is surprisingly large, and reflects the fact that magnetic diffusion along the mean field lines by low-$R_m$ Alfven waves is remarkably fast at $B$ = 30 Gauss.

Turning now to the other planets, we encounter the problem that, while $\Lambda$ and $R_\lambda$ can be estimated reasonably well, $\Pi_P$ is less well constrained. This means that the only property we can predict with confidence is $\hat{R}_m = u\delta_{\min}/\lambda$, which happens to be independent of $\Pi_P$. Predictions based on the mean axial field strength in the core, and an assumed rms value ten times larger, are given in Table 2, where the material properties are taken to be the same as those used in Table 1. While the estimates of $\hat{R}_m$ in Mercury, Earth and Saturn are reassuringly modest, those for Jupiter are sufficiently high to raise a question over the use of the low-$R_m$ approximation in calculating the induced emf in the columnar vortices.

| Planet | $\dfrac{\bar{B}_z/\sqrt{\rho\mu}}{\Omega R_C}$ | $R_\lambda = \dfrac{\Omega R_C^2}{\lambda}$ | $\hat{R}_m = \dfrac{u\delta_m}{\lambda} \sim \Lambda^{1/2}$ based on $\bar{B}_z$ | $\hat{R}_m = \dfrac{u\delta_m}{\lambda} \sim \Lambda^{1/2}$ based on $B_{rms} \sim 10\bar{B}_z$ |
|---|---|---|---|---|
| Mercury | $5.6\times10^{-6}$ | $4.0\times10^{6}$ | 0.011 | 0.11 |
| Earth | $13\times10^{-6}$ | $8.9\times10^{8}$ | 0.39 | 3.9 |
| Jupiter | $5.2\times10^{-6}$ | $1.3\times10^{11}$ | 1.9 | 19 |
| Saturn | $2.2\times10^{-6}$ | $3.4\times10^{10}$ | 0.41 | 4.1 |

Table 2. The order-of-magnitude prediction of $\hat{R}_m = u\delta_{\min}/\lambda$ using the scaling law $\hat{R}_m \sim \Lambda^{1/2}$. The pre-factor in this scaling relationship has been set to unity.

Finally we turn to the published numerical simulations. Here the definitions and notation adopted in most papers is somewhat different to ours. For example, both Ro and $\Pi_B$ (which is more commonly given the symbol Lo) use the annular gap, $D = R_C - R_i$, rather than $R_C$, as the characteristic length scale. Moreover

$$Ra_Q = \frac{1}{4\pi R_C R_i}\frac{g\beta}{\rho c_p}\frac{Q_T}{\Omega^3 D^2}$$

is often used instead of our $\Pi_p$, though for the Earth the two are related by $\Pi_p \approx 0.3 Ra_Q$.

When comparing our predictions with the simulations we encounter two difficulties. First, the magnetic Prandtl number in these simulations typically lies in the range $O(0.1) < P_m < O(10)$, as distinct from $\sim 10^{-6}$ in the Earth, reflecting the fact that the viscosity in the simulations is much too high. This is important because (5.15), which represents the energy balance $\bar{\mathrm{P}} = J^2/\rho\sigma$, holds only when the magnetic Prandtl number is small and the viscous dissipation weak. In short, our predictions hold only for small $P_m$. Second, the Rossby number in the simulations usually lies in the range $O(10^{-4}) < \mathrm{Ro} < O(10^{-1})$, as opposed to $\sim 10^{-6}$ in the Earth, reflecting the relatively low rotation rate in many of the numerical experiments. When $\mathrm{Ro} > 10^{-2}$ at the large scales, we would expect $u/\Omega\delta_{\min} > 1$ at the small scales, yet our entire model rests on the assumption that $u/\Omega\delta_{\min} < 1$, so that inertial waves can propagate freely at the small scales. For these reasons, we shall restrict ourselves to those numerical experiments in which $P_m < 1$ and $\mathrm{Ro} < 10^{-2}$.

Since the length scale $\delta_{\min}$ is rarely quoted in numerical datasets, we shall focus first on prediction (5.12), which does not involve $\delta_{\min}$:

$$\mathrm{Ro} \sim \frac{\Pi_P R_\lambda^{1/2}}{\Pi_B} \sim \frac{Ra_Q R_\lambda^{1/2}}{\mathrm{Lo}}. \qquad (5.18)$$

Note that (5.18) is independent of any assumption as to the physical processes that set $\delta_{\min}$, and so could, in principle, include cases where $\delta_{\min}$ is set by the viscous scale $\delta_{\min}/R_C \sim (\mathrm{Ek})^{1/3}$. We have used the dataset of Christensen & Aubert (2006), restricting ourselves to those simulations in which $P_m \le 1$ and $\mathrm{Ro} < 10^{-2}$. The results are shown in Figure 6, where the various numerical simulations are the points, and our prediction the straight line. The line corresponds to a pre-factor of 0.08, i.e. $\mathrm{Ro} = 0.08\, Ra_Q\, R_\lambda^{1/2}/\mathrm{Lo}$. The agreement seems not unreasonable, though of course we have excluded a significant part of the dataset in the comparison. The small value of the pre-factor in $\mathrm{Ro} = 0.08\, Ra_Q R_\lambda^{1/2}/\mathrm{Lo}$ might, in part, explain why we have overestimated Ro for the Earth.

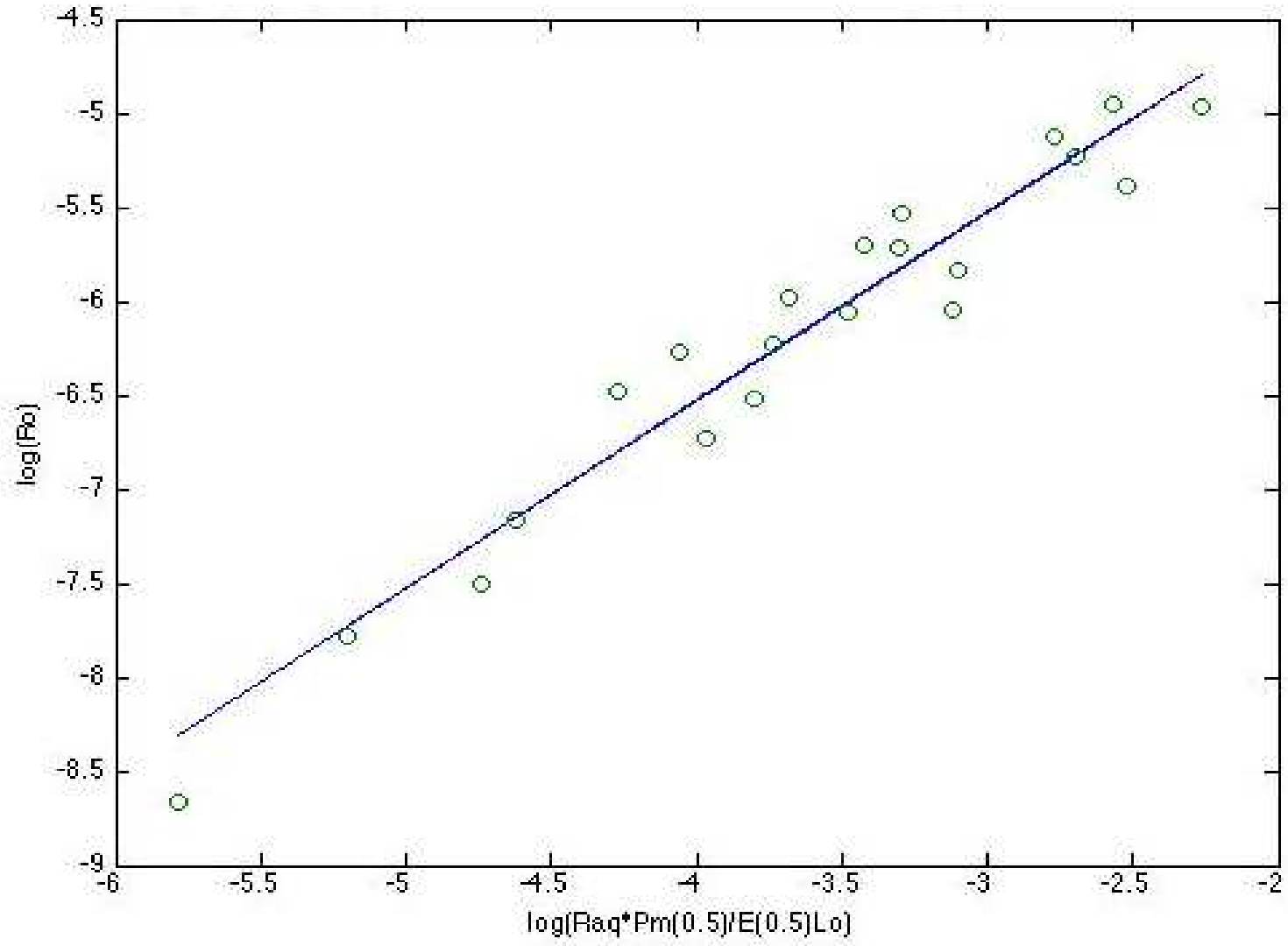

Figure 6. Comparison of prediction (5.18) with the numerical simulations of Christensen & Aubert (2006). The data is restricted to those simulations in which $P_m \le 1$ and $\mathrm{Ro} \le 0.008$. The *y* axis is ln Ro and the *x* axis $\ln Ra_Q R_\lambda^{1/2} \big/ \mathrm{Lo}$.

We close by returning to the observation that some authors have suggested that viscosity sets the scale for $\delta_{\min}$ in at least some of the numerical simulations (see, for example, King & Buffett, 2013), so it is of interest to return to (5.16) and (5.17). If we accept the suggestion that viscosity plays an important role in setting the scale for $\delta_{\min}$ in the numerical simulations (though presumably not in the planets), we would expect to observe

$$\mathrm{Ro} \sim \Pi_p^{1/2} \mathrm{Ek}^{-1/6}, \qquad (5.19)$$

$$\Pi_B \sim \Pi_p^{1/2} \mathrm{Ek}^{-1/3} P_m^{1/2}, \qquad (5.20)$$

in a regime in which inertial waves dominates over Ekman pumping in the production of helicity .

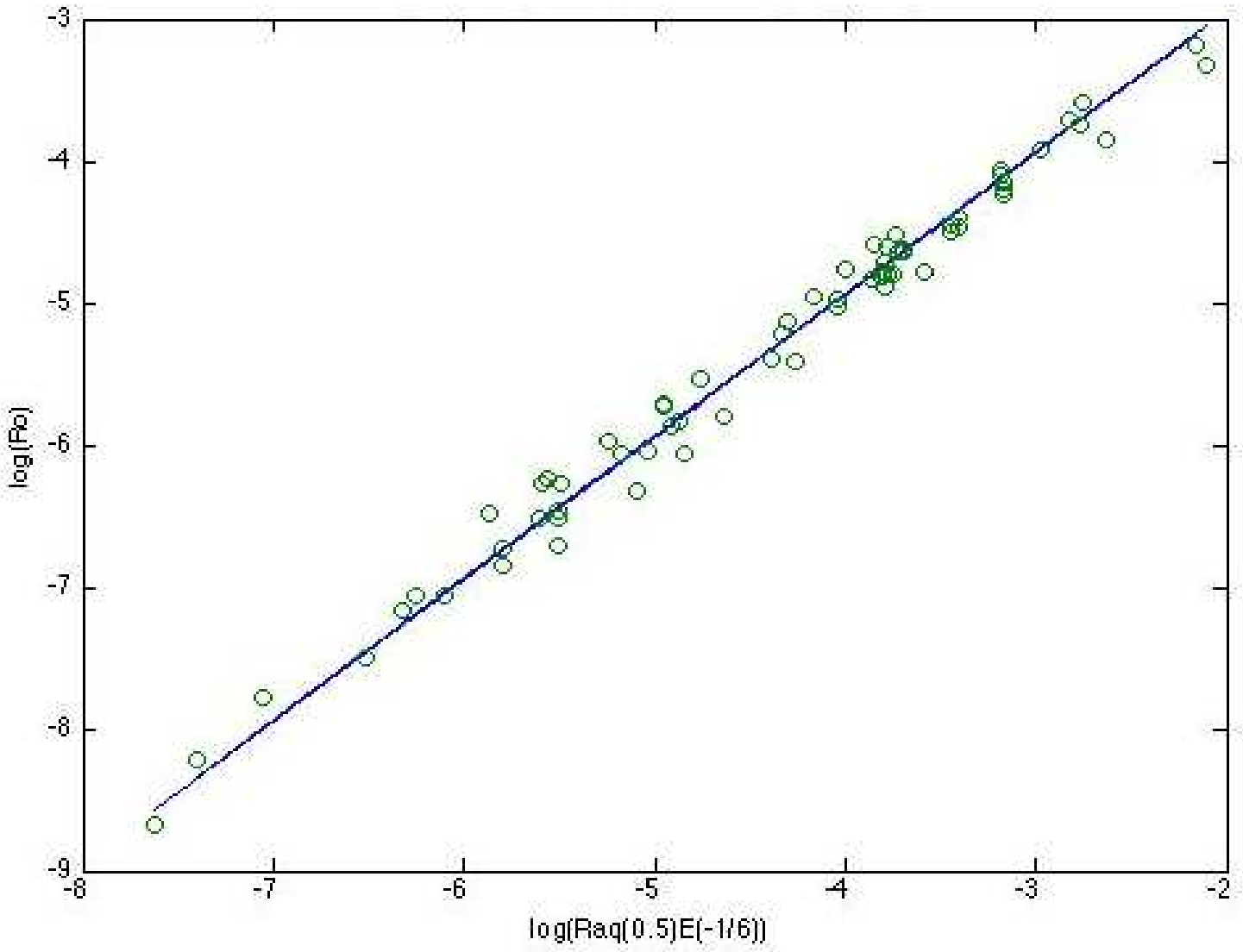

Figure 7. Comparison of prediction (5.19) with the numerical simulations of Christensen & Aubert (2006). The vertical axis is ln Ro and the horizontal axis $\ln Ra_Q^{1/2} \mathrm{Ek}^{-1/6}$. The full dataset is used.

The first of these scaling is put to the test in Figure 7, where once again we have used the dataset of Christensen & Aubert (2006). The results seem favourable, and it is striking that this scaling is a good match across the full dataset, which is to be expected since (5.19) rest simply on an assumed order of magnitude balance between viscous stresses, the Coriolis force, and buoyancy, so that the restrictions on $P_m$ and $\mathrm{Ro}$ do not apply. This lends tentative support to the assertion that viscosity plays a key role in setting $\delta_{\min}$ in the numerical simulations.

Next, prediction (5.20) is shown in Figure 8. In the top panel the full dataset is used and the results are not particularly favourable, suggesting different behaviour at low and high Lo (which corresponds to low and high Ro). However, in the lower panel the dataset is restricted to simulations in which $P_m \leq 1$ and $\mathrm{Ro} < 10^{-2}$, which is the regime in which our theory applies. Here the comparison is more favourable. (A similar result may be obtained by filtering on $Ra_Q < 0.1\mathrm{Ek}$, corresponding roughly to $u/\Omega\delta_{\min} < 1$.) One interpretation of the top panel is that the point at which the data starts to fall below our prediction corresponds to $u/\Omega\delta_{\min} \sim 1$, i.e. $\overline{\mathrm{P}}/\Omega^2\nu \sim 1$. As the rotation is decreased further, i.e. we move further to the right, the inertial waves become sparser and the task of producing helicity falls increasingly to Ekman pumping, with a corresponding drop in magnetic energy.

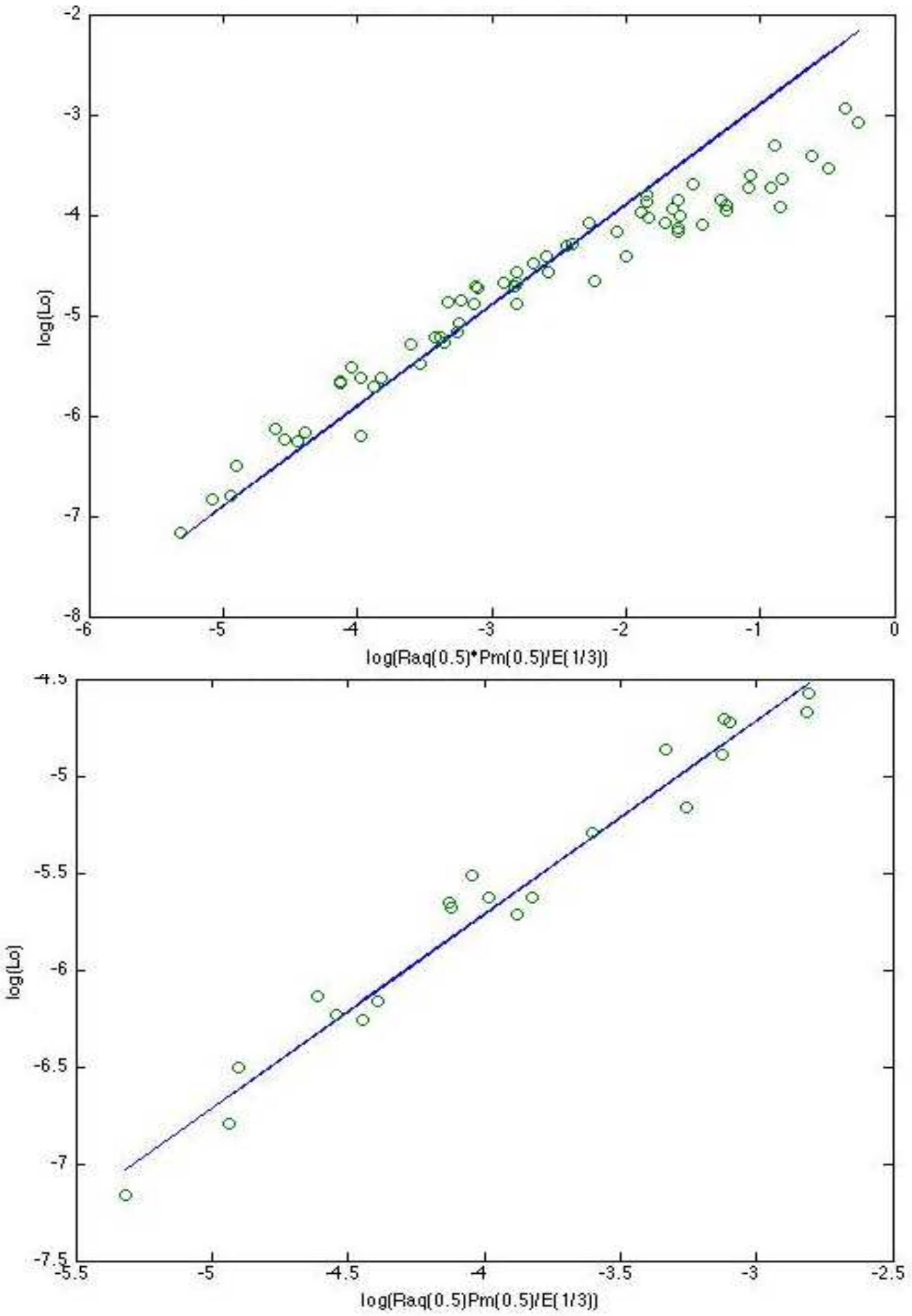

Figure 8. Comparison of prediction (5.20) with the numerical simulations of Christensen & Aubert (2006). The vertical axis is ln $\Pi_B$ and the horizontal axis ln $Ra_Q^{1/2} Pm^{1/2} \big/ \mathrm{Ek}^{1/3}$. In the top panel the full dataset is used, and in the lower panel the dataset is restricted to simulations in which $P_m \leq 1$ and $\mathrm{Ro} \leq 0.008$.

We might tentatively conclude that (5.19) and (5.20) are both reasonably well supported by the data, at least for the restricted dataset. Let us now rewrite (5.19) and (5.20) as

$$\mathrm{Ro} \sim \Pi_p^{4/9} P_m^{-1/6} \left( \frac{\overline{\mathrm{P}}^{1/3} R_C^{4/3}}{\lambda} \right)^{1/6}, \qquad (5.21)$$

$$\Pi_B \sim \Pi_p^{7/18} P_m^{1/6} \left( \frac{\overline{\mathrm{P}}^{1/3} R_C^{4/3}}{\lambda} \right)^{1/3}. \qquad (5.22)$$

It turns out that $\left(\overline{\mathrm{P}}^{1/3} R_C^{4/3} / \lambda\right)^{1/3}$ does not vary much in most numerical datasets; for example, it takes a value of $10 \pm 5.6$ in the full dataset of Christensen & Aubert (2006). So, for these kinds of limited datasets, (5.21) and (5.22) would not look very different to $\mathrm{Ro} \sim \Pi_p^{4/9} P_m^{-1/6}$ and $\Pi_B \sim \Pi_p^{7/18} P_m^{1/6}$. In this respect it is interesting to note that Stelzer & Jackson (2013) report the empirical relationships $\mathrm{Ro} \sim \Pi_p^{0.44} P_m^{-0.13}$ and $\Pi_B \sim \Pi_p^{0.31} P_m^{0.16}$ for their numerical datasets, and indeed empirical scalings close to $\mathrm{Ro} \sim \Pi_p^{0.44}$ and $\Pi_B \sim \Pi_p^{0.33}$ have been reported by a number of authors (see, for example, Davidson, 2013b, for a summary).

## 6. **Conclusions**

In many mildly supercritical numerical simulations the dynamo mechanism is viscously driven, taking the form of helical Ekman pumping within columnar convection rolls. This Ekman pumping arises from the viscous interaction of the convection columns with the mantle. However, given the similarity of the external magnetic fields observed in the terrestrial planets and gas giants (which have no mantle), and the extremely small value of the Ekman number in all such cases, it seems natural to suppose that the mechanism of helicity generation in the planets is independent of viscosity and insensitive to mechanical boundary conditions. In this paper we have proposed that helicity in the core of the Earth arises from the spontaneous emission of inertial waves, driven by the equatorial heat flux in the outer core. We have demonstrated that such waves produce the required helicity distribution outside the tangent cylinder (negative in the north and positive in the south), and have shown that these waves inevitably propagate along the axis of the columnar vortices, and indeed they are the very mechanism by which the columnar vortices first form and then subsequently evolve. Moreover, we have calculated the emf induced by such axially propagating inertial waves and shown that, in principle, this emf is sufficient to support a self-sustaining dynamo of the $\alpha^2$ type. Finally, we have derived the scaling laws for this kind of inertial-wave planetary dynamo, which operates in the regime of: (i) $u\delta_{\min}/\lambda < 1$; (ii) $u/\Omega\delta_{\min} < 1$; and (iii) small $P_m$. These predictions compare favourably with those numerical simulations which fall into the appropriate regime and also with what little we know about the Earth's core.

The requirement that $u/\Omega\delta_{\min} < 1$, which is almost certainly satisfied in the planets, holds in only some of the numerical experiments. In those simulations which fail to satisfy $u/\Omega\delta_{\min} < 1$, we would expect either reduced helicity generation via Ekman pumping, or else negligible helicity. In the latter case, large-scale dynamo action is difficult to achieve, with magnetic energy generation being confined to the small scales. This might explain the observation of Christensen & Aubert (2006) that quasi-steady dipolar dynamos are difficult to realise for $u/\Omega\ell < 0.1$, where their $\ell$ is a length scale whose definition makes it somewhat larger than our $\delta_{\min}$.

The author is pleased to acknowledge the help of Avishek Ranjan, who performed the simulation shown in Figure 5, and Chris Walsh, who produced the figures in §5.